\documentclass[preprint,12pt]{elsarticle}

\usepackage{bm}
\usepackage{amsmath}
\usepackage{amssymb}
\usepackage{mathtools}
\usepackage{siunitx}
\usepackage{fancyref}
\usepackage{pgfplots}
\usepackage{pgfplotstable}
\usepgfplotslibrary{external}

\usepackage{subcaption}
\usepackage{todonotes}
\usepackage{eurosym}
\usepackage{multirow}
\usepackage{tabularx}
\usepackage{booktabs}
\usepackage{subcaption}
\usepackage{mwe}
\usepackage[nomessages]{fp}
\usepackage{booktabs}
\usepackage{tabularx}

\usepackage[acronym]{glossaries}

\usepackage{layout}
\usepackage{printlen}
\usepackage{hyperref}

\usetikzlibrary{math} 

\DeclareSIUnit{\ct}{ct}
\DeclareSIUnit\year{yr}

\journal{Applied Energy}

\newacronym{nlp}{NLP}{Non Linear Problem}
\newacronym{milp}{MILP}{Mixed Integer Linear Program}
\newacronym{minlp}{MINLP}{Mixed Integer Non-Linear Program}
\newacronym{dhn}{DHN}{District Heating Networks}
\newacronym{gis}{GIS}{Geographic Information System}
\newacronym{chp}{CHP}{Combined Heat and Power}
\newacronym{4gdh}{4GDH}{4th Generation District Heating}
\newacronym{capex}{CAPEX}{Capital Expense}
\newacronym{opex}{OPEX}{Operational Expense}

\makeatletter
\renewcommand{\todo}[2][]{\tikzexternaldisable\@todo[#1]{#2}\tikzexternalenable}
\makeatother

\begin{document}

\setlength{\parindent}{0cm}

\DeclareSIUnit{\sieuro}{\mbox{\euro}}

\begin{frontmatter}

\title{A Multi-Period Topology and Design Optimization Approach for District Heating Networks}

\author{Yannick Wack \fnref{label1,label2,label3}}
\ead{yannick.wack@kuleuven.be}
\author{Martin Sollich \fnref{label1,label3}}
\author{Robbe Salenbien \fnref{label2,label3}}
\author{Jan Diriken \fnref{label2,label3}}
\author{Martine Baelmans \fnref{label1,label3}}
\author{Maarten Blommaert \fnref{label1,label3}}

\fntext[label1]{Department of Mechanical Engineering, KU Leuven, Celestijnenlaan 300 box 2421, 3001 Leuven, Belgium}
\fntext[label2]{Flemish Institute for Technological Research (VITO), Boeretang 200, 2400 Mol, Belgium}
\fntext[label3]{EnergyVille, Thor Park, Poort Genk 8310, 3600 Genk, Belgium}

\begin{abstract}
The transition to 4th generation district heating creates a growing need for scalable, automated design tools that accurately capture the spatial and temporal details of heating network operation. This paper presents an automated design approach for the optimal design of district heating networks that combines scalable density-based topology optimization with a multi-period approach. In this way, temporal variations in demand, supply, and heat losses can be taken into account while optimizing the network design based on a nonlinear physics model. The transition of the automated design approach from worst-case to multi-period shows a design progression from separate branched networks to a single integrated meshed network topology connecting all producers. These integrated topologies emerge without imposing such structures a priori. They increase network connectivity, and allow for more flexible shifting of heat loads between different producers and heat consumers, resulting in more cost-effective use of heat. In a case study, this integrated design resulted in an increase in waste heat share of 42.8 \% and a subsequent reduction in project cost of 17.9 \%. We show how producer unavailability can be accounted for in the automated design at the cost of a 3.1 \% increase in the cost of backup capacity. The resulting optimized network designs of this approach connect multiple low temperature heat sources in a single integrated network achieving high waste heat utilization and redundancy, highlighting the applicability of the approach to next-generation district heating networks.
\end{abstract}


\end{frontmatter}

\setcounter{footnote}{0}

\newcommand{\tp}[1]{#1^{\intercal}} 			
\DeclarePairedDelimiter\abs{\lvert}{\rvert} 	
\newcommand{\card}[1]{\lvert#1\rvert}
\newcommand{\infNorm}[1]{\|#1\|_{\infty}}
\newcommand{\Real}[1]{\mathbb{R}^{#1}}
\newcommand{\ve}[1]{\bm{#1}} 	

\newcommand{\n}{n}
\newcommand{\nNodes}{\n_{\mathrm{N}}}
\newcommand{\nEdges}{\n_{\mathrm{E}}}
\newcommand{\nElements}{\n_{\textrm{spat}}}
\newcommand{\nModels}{\n_{\model}}
\newcommand{\ntechConst}{\n_{\mathrm{tech}}}

\newcommand{\npipes}{\n_{\pipe}}
\newcommand{\nCon}{\n_{\con}} 
\newcommand{\nPro}{\n_{\pro}} 
\newcommand{\nRad}{\n_{\rad}} 
\newcommand{\nByp}{\n_{\byp}} 

\newcommand{\nperiods}{\n_{\textrm{period}}}
\newcommand{\timeSlices}{\nperiods}

\newcommand{\defnNodes}{\nNodes = \card{\setNodes}}
\newcommand{\defnEdges}{\nEdges = \card{\setEdges}}
\newcommand{\defnPeriods}{\nperiods \in \mathbb{N}}
\newcommand{\defnpipes}{\npipes = \card{\Epipe}}
\newcommand{\defnCon}{\nCon = \card{\Econ}} 
\newcommand{\defnPro}{\nPro = \card{\Epro}} 
\newcommand{\defnRad}{\nRad = \card{\Erad}} 
\newcommand{\defnByp}{\nByp = \card{\Ebyp}} 
\newcommand{\defntechConst}{\ntechConst\in\mathbb{N}_1}
\newcommand{\defnElements}{\nElements = \nEdges+\nNodes}

\newcommand{\equalNew}{\tilde{\equalCon}}
\newcommand{\inEqualNew}{\tilde{\inEqualCon}}
\newcommand{\designVarNew}{\tilde{\designVar}}

\newcommand{\radFlowNew}{\tilde{\radValve}}
\newcommand{\prodInputNew}{\tilde{\prodInput}}
\newcommand{\equalConModel}{\equalCon_\mathrm{m}}
\newcommand{\inEqualModel}{\equalCon_\mathrm{s}}
\newcommand{\stateVarModel}{\stateVar_\mathrm{m}}
\newcommand{\stateVarIneq}{\stateVar_\mathrm{s}}

\newcommand{\ALagrangian}{\mathcal{L}}
\newcommand{\LagMultis}{\lambda}
\newcommand{\LagPen}{\mu}
\newcommand{\slack}{s}
\newcommand{\equalConState}{g}

\newcommand{\costFull}{\mathcal{J}}
\newcommand{\costi}[2][]{\cost_{\mathrm{#2}#1}}
\newcommand{\subCAPEX}{CAP}
\newcommand{\subOPEX}{OP}
\newcommand{\subPipe}{pipe}
\newcommand{\subHeat}{h}
\newcommand{\subPump}{p}
\newcommand{\subRev}{rev}

\newcommand{\periodWeights}{w}

\newcommand{\Jpipepol}{\kappa}
\newcommand{\Jpipesmooth}{\xi}
\newcommand{\cPipe}{\npvCost_{mathrm{\subPipe}}}

\newcommand{\cHeatOPEX}{\npvCost_{\mathrm{hO}}}
\newcommand{\cHeatOPEXi}[1]{\npvCost_{\mathrm{hO},#1}}

\newcommand{\pumpEff}{\efficiency_{\mathrm{pump}}}
\newcommand{\cPumpOPEX}{\npvCost_{\mathrm{pO}}}
\newcommand{\cPumpOPEXi}[1]{\npvCost_{\mathrm{pO},#1}}
\newcommand{\cPumpCAPEX}{\npvCost_{\mathrm{pC}}}
\newcommand{\cPumpCAPEXi}[1]{\npvCost_{\mathrm{pC},#1}}

\newcommand{\cRev}{\npvCost_{\mathrm{r}}}
\newcommand{\cRevi}[1]{\npvCost_{\mathrm{r},#1}}
\newcommand{\flow}{q}
\newcommand{\pressure}{p}
\newcommand{\temperature}{T}

\newcommand{\Reynolds}{Re}      
\newcommand{\density}{\rho}     
\newcommand{\viscosity}{\mu}    
\newcommand{\spHeatCap}{c_{\mathrm{p}}}

\newcommand{\TOutside}{\temperature_\infty}
\newcommand{\dTinf}{\theta}                 
\newcommand{\heat}{Q}

\newcommand{\length}{L} 
\newcommand{\diameter}{d}
\newcommand{\diameterMin}{\diameter_{\mathrm{min}}}
\newcommand{\diameterDiscrete}{D}
\newcommand{\volume}{V}

\newcommand{\rough}{\epsilon} 
\newcommand{\ratioInsul}{r} 
\newcommand{\condInsul}{\lambda_{\mathrm{i}}} 
\newcommand{\condGround}{\lambda_{\mathrm{g}}} 

\newcommand{\subScriptOuterD}{o}  
\newcommand{\depthPipe}{h} 

\newcommand{\hydrR}{R}  
\newcommand{\frictionfactor}{f} 	
\newcommand{\thermR}{U} 

\newcommand{\inflowID}{a}    
\newcommand{\outflowID}{b}

\newcommand{\valveRhydr}{\zeta}   

\newcommand{\bypValve}{\alpha}     

\newcommand{\lmtd}{LMTD}
\newcommand{\heaterCoef}{\xi}	
\newcommand{\heaterExp}{n} 
\newcommand{\THouse}{\dTinf_{\textrm{house}}}

\newcommand{\Qdemand}[1][]{\heat_{\mathrm{d}#1}}
\newcommand{\Qdemandi}[1]{Q_{\mathrm{d},#1}}
\newcommand{\DemandSatisfaction}{S}

\newcommand{\prodInputi}[1]{\prodInput_{#1}} 	

\newcommand{\prodTemp}{\Theta}
\newcommand{\prodTempv}{\ve{\Theta}}
\newcommand{\prodTempi}[1]{\Theta_{#1}} 

\newcommand{\existance}{\phi}
\newcommand{\exPipe}{\phi}
\newcommand{\exProducer}{\omega}

\newcommand{\heatInstalled}{H}
\newcommand{\flowVelocity}{v}
\newcommand{\massFlow}{m}
\newcommand{\bigM}{\mathcal{M}}
\newcommand{\conSimpleHX}{c}

\newcommand{\segment}{s}

\newcommand{\gams}{fMINLP}
\newcommand{\pathopt}{pNLP}

\newcommand{\walltime}{w}
\newcommand{\walltimeGAMS}{\walltime_{\mathrm{\gams{}}}}
\newcommand{\walltimePATHOPT}{\walltime_{\mathrm{\pathopt{}}}}

\newcommand{\numberPipes}{n}
\newcommand{\pen}{\xi} 
\newcommand{\penDirection}{a}
\newcommand{\penDiameter}{\bar{\diameter}}
\newcommand{\setDiameters}{\mathcal{S}}
\newcommand{\relu}{f}


\newcommand{\defSetPipes}{\forall \gi\gj \in \Epipe}
\newcommand{\defSetRad}{\forall \gi\gj \in \Econ}
\newcommand{\defSetByp}{\forall \gi\gj \in \Ebyp}
\newcommand{\defSetCon}{\forall \gi\gj \in \Econ}
\newcommand{\defSetProd}{\forall \gi\gj \in \Epro}

\newcommand{\defInflow}{\inflowID=(\gi,n) \in \setEdges}
\newcommand{\defOutflow}{\outflowID=(n,\gj) \in \setEdges}

\newcommand{\topVarBoxConstraints}[1]{\diameterDiscrete_{0}\leq#1\leq\diameterDiscrete_{N}}
\newcommand{\designVarUpper}{\designVar_{\mathrm{up}}}
\newcommand{\designVarLower}{\designVar_{\mathrm{low}}}
\newcommand{\opVarBoxConstraints}[1]{\designVarLower\leq #1 \leq \designVarUpper}
\newcommand{\defFlow}{\ve{\flow}_{\timeVar} \in \Real{\nEdges}}
\newcommand{\defPressure}{\ve{\pressure}_{\timeVar} \in \Real{\nNodes}}
\newcommand{\defTemp}{\ve{\dTinf}_{\timeVar} \in \Real{\nEdges+\nNodes} }

\newcommand{\defDesignVar}{\designVar = \tp{\left[\tp{\topVar},\tp{\ve{\capVar}},\tp{\ve{\radValve}},\tp{\ve{\prodInput}}\right]}}
\newcommand{\defTopVar}{\topVar\in\Real{\npipes}}

\newcommand{\defModelConstraints}{\equalCon_\timeVar\left( \designVar,\stateVar_\timeVar\right)}

\newcommand{\interestRate}{r_{\mathrm{in}}}
\newcommand{\maxPressure}{\Delta p_\mathrm{max}}
\newcommand{\peakPeriod}{t_{\textrm{peak}}}
\newcommand{\setPeriods}{\Upsilon}

\section{Introduction}
\gls{dhn}s are a crucial factor in the energy transition towards carbon neutral space heating \cite{OECD/IEA2019}. Modern \gls{dhn}s provide an energy infrastructure that transports hot water in insulated pipes and connects a wide range of renewable and low-carbon heat sources, such as geothermal, waste heat, or biomass, to the residential and commercial demand of districts or entire cities \cite{Lund2018}. In particular, modern \gls{4gdh} has grown in popularity as it helps to reduce $\mathrm{CO}_2$ emissions and increase the overall efficiency of energy use and increase energy independence in the heating sector. 

The design of these modern DHNs is challenging due to the need to connect heat sources of different temperature levels (e.g. high temperature back-up boilers and low temperature waste heat sources) with a variety of different heat demands ranging from large commercial buildings to private homes with varying levels of renovation. The \gls{capex} intensity of \gls{dhn}s, especially the ground work and piping, is often the bottleneck for the feasibility of a development project. Therefore, it is important to optimize key network features such as network topology (routing), pipe sizes, and heat producer capacities early in the design phase. Here, optimization and automated design tools can assist in increasing energy efficiency and reducing network costs. They can also assist in assessing the viability of heating network projects and reduce uncertainty for investors, ultimately helping to increase the adoption of this technology.

Therefore, an increasing amount of research has focused in recent years on optimizing the design of DHNs. The design of modern DHNs often requires a detailed understanding of the future operation of the network. Optimizing the topology of a network based on a nonlinear physics model that simulates this future operation constitutes a \gls{minlp}. \gls{minlp}s are notoriously difficult to solve, and the literature on approaches to solving the \gls{minlp} routing problem for DHNs can be divided into four categories. Directly solving the \gls{minlp} using heuristic approaches such as genetic algorithms as done by Li and Svendsen \cite{Li2013} or Egberts et al. \cite{Egberts2020a} or a particle swarm algorithm as done by Allen et al. \cite{Allen2022}. Others solve the \gls{minlp} directly using combinatorial optimization, such as Mertz et al. \cite{mertz2016minlp} or Marty et al. \cite{marty2018simultaneous}. This method is challenging and can become intractable for large problems \cite{Wack2023}. As a result, the topology optimization problem of DHNs is often linearized, allowing it to be solved efficiently using \gls{milp} solvers. Examples include the work of Weinand et al. \cite{Weinand2019}, Resimont et al. \cite{Resimont2021}, and Neri et al. \cite{Neri2022}. The interested reader is referred to Wack et al. \cite{Wack2023} for a more comprehensive review of topology optimization approaches for \gls{dhn}s.

The trajectory of modern DHNs to push for ever lower network temperatures and to include multiple heat sources at different temperature levels increases the importance of a nonlinear physics model that can accurately model heat losses and account for feasible temperature levels at the different supply and demand locations. Linearized approaches often can't solve the original physics-based problem and have to rely on a priori assumptions, such as a fixed flow direction. An alternative way to improve tractability is to use a density-based topology optimization approach that preserves the nonlinear representation of the network physics. By initially relaxing the integer constraint on pipe placement, it is possible to efficiently solve a nonlinear programming (NLP) problem and still achieve a near-discrete design. This method has been formally introduced by Blommaert et al. \cite{Blommaert2020a} and Wack et al. \cite{Wack2022a}, allowing them to optimize the topology of a \gls{dhn} based on a nonlinear thermal-hydraulic model, and ensuring a discrete network topology through a density-based formulation. The method remains tractable for large-scale problems \cite{Wack2023}. Previously, Pizzolato et al. \cite{Pizzolato2018} used a similar method to robustly optimize the topology of a DHN based on a hydraulic network model.  However, this density-based approach for DHNs was previously limited to a single steady-state period and only allowed for worst-case or average scenario optimization. Therefore, it cannot account for temporal variations in e.g. heat demand and supply, which is particularly important for modern \gls{4gdh} that integrate renewable heat sources whose supply is often intermittent.

\subsection*{Time dependent optimal design of district heating networks}

Accounting for temporal variations in heat supply and demand in the design optimization of \gls{dhn}s requires a time resolution. To avoid the computational complexity of modeling the transient behavior of the network with dynamic hydraulic and/or thermal models, multi-period approaches are used. This approach assumes hydraulic and thermal equilibrium and is often used when the application permits. These methods typically define a small subset of time periods with the goal of representing the full time resolution of the optimization problem. 

An early example is S\"oderman \cite{Soderman2007}, who optimize the routing and positioning of production and storage sites in a district cooling system using an \gls{milp} based on a multi-period approach. They represent a full year's cooling demand with 8 periods for day and night in each season. Weber and Shah \cite{Weber2011} present an optimization tool that optimizes the technology mix of a district energy system based on a selection of 6 periods for 3 representative days for summer, mid-season and winter. Li and Svendsen \cite{Li2013a} optimized the topology of a \gls{dhn} dividing the annual heat load into 8 representative periods. Bracco et al. \cite{Bracco2013} optimize the design and operation of a heat distribution system with distributed \gls{chp} units. This optimization considers a time horizon of four representative days, including winter, spring, summer, and fall. Haikarainen et al. \cite{Haikarainen2014} optimize a \gls{dhn} focusing on the technology and location of the energy supplier, the network topology, and the operation of heat storage. They assume typical consumption profiles for heat demand and divide them into two-monthly intervals, with one day and one night period for each interval, for a total of 12 periods considered for this problem. 

In a more systematic approach, Fazlollahi et al. \cite{Fazlollahi2014} reduce energy demand profiles into typical periods, trying to adequately preserve the characteristics of the annual profile using a k-means based clustering method. They then use this approach to aggregate energy demand profiles, solar irradiation, and electricity prices of a typical year into 8 typical periods to optimize the design and operation strategy of a district energy system. Morvaj et al. \cite{Morvaj2016} optimize the operation and design of a \gls{dhn} to minimize cost and carbon emissions. They aggregate heat demand time series into an average 24-hour demand curve for each month of the year. To determine the optimal capacity and operation of a combined cooling, heating, and power system, Ameri and Besharati \cite{Ameri2016} optimize a heating and cooling network. This optimization is based on the hourly heating, cooling, and electricity demand for a typical summer and winter day. To optimize the mix of technologies and marginal expansion of \gls{dhn}, Delangle et al. \cite{Delangle2017} consider a large set of 6912 time periods representing hourly demand data for 2 typical days per month over a span of 12 years. To optimize the design and operation of a long-distance heat transport system, Hirsch et al. \cite{Hirsch2018} divide the temporal resolution of the optimization problem into 3 long horizon periods and 7 short horizon periods. Focusing on generation, conversion, transport, and storage technologies, Samsatli and Samsatli \cite{Samsatli2018} optimize an urban energy system based on 6 representative periods for three typical days in winter, summer, and midseason. Weinand et al. \cite{Weinand2019} optimize the topology and location of heating plants of a \gls{dhn}. This optimization problem enforces hourly heat supply to consumers for a full year, considering 8760 periods. 

Again shifting to a more structured approach of time aggregation, Van der Heijde et al. \cite{VanderHeijde2019c} optimize the design of a \gls{dhn} focusing on the size and location of solar thermal collectors, seasonal thermal storage, and excess heat. They use a representative days method \cite{VanderHeijde2019b} based on Poncelet et al. \cite{Poncelet2017} to aggregate annual heat demand, solar radiation, ambient temperature, and hourly electricity price into 12 representative days. The same representative day approach was used by Resimont et al. \cite{Resimont2021} to optimize an urban heating network, aggregating heat demand data into 144 representative periods. Neri et al. \cite{Neri2022} determine the optimal topology, pipe diameter, and set of consumers connected to a district cooling network. They aggregate heat demand profiles into three 8-hour averages for morning, afternoon, and night. Wirtz et al. \cite{Wirtz2023} optimally select and size energy conversion units in buildings and energy hubs connected to a bidirectional low-temperature network. They use k-medoids clustering to aggregate a full year's time series into 16 representative days. To improve the integration of solar thermal in \gls{dhn}s, Delubac et al. \cite{Delubac2023}, optimally size and operate thermal storage together with different heating plants. They use a k-medoids algorithm to select 144 representative days for optimization. None of these approaches to date allows for nonlinear physics-based and scalable network topology and design optimization that accounts for temporal changes.

\subsection*{Time aggregation methods}

These examples of multi-period optimal design of \gls{dhn}s show the great diversity both in the number of considered periods and in the approaches to selecting these representative periods. The number of representative periods selected is a trade-off between computational complexity and accuracy of the optimization problem. The computational efficiency of linear approaches often leaves room for a more extensive temporal scope, like Delangle et al. \cite{Delangle2017}, who considered more than 6000 periods. The more extensive and detailed the optimization problem becomes, the more important an efficient choice of the number of periods becomes. Many nonlinear approaches therefore consider only a single period or scenario, such as Mertz. et al. \cite{mertz2017minlp} or Roland and Schmidt \cite{Roland2020}. However, a single-period approach is often insufficient to accurately account for the temporal resolution of, for example, heat demand, outdoor temperatures, or production unavailability. 

As evident from these examples in the literature, the number of periods considered in an optimization must be carefully chosen to ensure an efficient and feasible optimized network design while maintaining computational tractability for large \gls{dhn} projects. To manage and reduce the computational complexity introduced by the temporal scope of a multi-period approach, time aggregation methods are often used. Early on, this aggregation was often done heuristically, choosing representative periods based on patterns in annual data such as day and night cycles (e.g. S\"oderman \cite{Soderman2007}), the four seasons (e.g. Weber and Shah \cite{Weber2011}), or the months of the year (e.g. Haikarainen et al. \cite{Haikarainen2014}). In recent years, this selection method has shifted towards more structured approaches to aggregate a high-resolution time series into representative periods. Van der Heijde et al. \cite{VanderHeijde2019b} developed a representative days method based on Poncelet et al. \cite{Poncelet2017} formulated as an \gls{milp} that minimizes the difference between the duration curves of a full year and a representative year for different time series. In this way, an a priori defined number of representative days is selected. This method also recovers the chronology of the selected days to be able to use them for e.g. thermal storage optimization.

Similarly, with the goal of aggregating time series for general optimal energy system design, Kotzur et al. \cite{Kotzur2018} propose and compare four different aggregation methods. They propose an aggregation method that normalizes time series, then aggregates the time series using averaging, k-means, k-medoids, or hierarchical clustering, and finally adds extreme periods to the set of representative periods.  A comparison is made for different energy system design problems, including a \gls{chp}-based system, a residential system based on a heat pump and photovoltaics, and an island system with a high share of renewable energy \cite{Kotzur2018}. They conclude that the aggregation method has little impact on the optimal design, with a preference for the k-medoids and hierarchical clustering methods \cite{Kotzur2018}. The study further shows that the trade-off between the depth of time series reduction and the modeling error is highly dependent on the system configuration, with centralized supply resources being able to be represented with a few typical days \cite{Kotzur2018}. Therefore, the impact of time series aggregation should be evaluated separately for each problem \cite{Kotzur2018}. 

\subsection*{Goal and scope of this paper}

The transition to \gls{4gdh} creates a growing need for scalable, automated design tools that accurately capture the spatial and temporal details in the operation of these modern networks. Therefore, the goal of this paper is to present an optimization approach for the optimal design of DHNs that combines scalable density-based topology optimization with a multi-period approach to allow resolving temporal variations of key parameters including heat demand, outdoor temperature, and unavailability of producer supply. The ability to optimize the network design based on a nonlinear physics model and to account for spatial and temporal changes in heat loads throughout the network enables the applicability of this scalable automated design approach to next-generation \gls{dhn}s. 

The contributions of this paper are organized as follows. First, the multi-period topology optimization problem is defined for \gls{dhn}s. Then, the methods for efficiently solving this problem are presented, and a time aggregation method is used to manage the computational complexity of the multi-period formulation. Finally, a case study is presented to demonstrate the potential of an automated multi-period design tool based on density-based topology optimization for the design of modern \gls{dhn}s. The integral design changes from worst-case to multi-period optimization on the optimal topology and design of the network are studied in detail, and the potential for increasing redundancy against producer unavailability is shown. 

\section{A multi-period topology optimization framework}\label{sec:OptProblem}
\newcommand{\gEdge}[3]{#1_{#2#3}}   
\newcommand{\gNode}[2]{#1_{#2}}     

\newcommand{\dirGraph}{G}
\newcommand{\setNodes}{N}
\newcommand{\setEdges}{E}

\newcommand{\pro}{\mathrm{pr}}
\newcommand{\con}{\mathrm{con}}
\newcommand{\rad}{\mathrm{hs}}
\newcommand{\byp}{\mathrm{bp}}
\newcommand{\jun}{\mathrm{jun}}
\newcommand{\pipe}{\mathrm{pipe}}
\newcommand{\operation}{\mathrm{op}}
\newcommand{\Npro}{\setNodes_\pro}
\newcommand{\Ncon}{\setNodes_\con}
\newcommand{\NconF}{\setNodes_{\con,\mathrm{f}}}
\newcommand{\NconR}{\setNodes_{\con,\mathrm{r}}}
\newcommand{\Njun}{\setNodes_\jun}
\newcommand{\NproF}{\setNodes_{\pro,\mathrm{f}}}
\newcommand{\NproR}{\setNodes_{\pro,\mathrm{r}}}

\newcommand{\EF}{\setEdges_{\mathrm{f}}}
\newcommand{\Epro}{\setEdges_\pro}
\newcommand{\EproF}{\setEdges_{\pro,\mathrm{f}}}
\newcommand{\EproR}{\setEdges_{\pro,\mathrm{r}}}

\newcommand{\Econ}{\setEdges_\con}
\newcommand{\Erad}{\setEdges_\rad}
\newcommand{\Ebyp}{\setEdges_\byp}
\newcommand{\Epipe}{\setEdges_\pipe}
\newcommand{\EpipeF}{\setEdges_{\pipe,\mathrm{f}}}
\newcommand{\EpipeR}{\setEdges_{\pipe,\mathrm{r}}}
\newcommand{\Eop}{\setEdges_\operation}

\newcommand{\gi}{i}
\newcommand{\gj}{j}

\newcommand{\giNode}[1]{\gNode{#1}{\gi}}
\newcommand{\gjNode}[1]{\gNode{#1}{\gj}}
\newcommand{\gijEdge}[1]{\gEdge{#1}{\gi}{\gj}}
\newcommand{\gitNode}[1]{\gNode{#1}{\gi,\timeVar}}
\newcommand{\gjtNode}[1]{\gNode{#1}{\gj,\timeVar}}
\newcommand{\gijtEdge}[1]{\gEdge{#1}{\gi}{\gj,\timeVar}}

\newcommand{\cost}{J}
\newcommand{\equalCon}{\ve{c}}
\newcommand{\inEqualConSkalar}{\ve{h}}
\newcommand{\inEqualConSkal}{h}
\newcommand{\inEqualCon}{\inEqualConSkalar}
\newcommand{\designVarskal}{\varphi}
\newcommand{\designVar}{\ve{\designVarskal}}
\newcommand{\designVarTimeInvariantSkalar}{\phi}
\newcommand{\designVarTimeInvariant}{\ve{\designVarTimeInvariantSkalar}}
\newcommand{\designVarTimeDependentSkalar}{\bar{\phi}}
\newcommand{\designVarTimeDependent}{\ve{\designVarTimeDependentSkalar}}
\newcommand{\stateVarSkal}{x}
\newcommand{\stateVar}{\ve{\stateVarSkal}}
\newcommand{\topVarSkalar}{d}
\newcommand{\topVar}{\ve{\topVarSkalar}}

\newcommand{\capVar}{\phi} 	
\newcommand{\prodInput}{\gamma} 	
\newcommand{\radValve}{\alpha}    
\newcommand{\timeVar}{t}
\newcommand{\techCon}{\ve{\inEqualConSkal}_{\mathrm{tech}}}
\newcommand{\defStateConstraints}{\techCon(\designVar,\stateVar)}

\newcommand{\nDiscretePipes}{\n_{\mathrm{D}}}
\newcommand{\setDefDiscreteDiameters}{\{\diameterDiscrete_{0},\dots,\diameterDiscrete_{\nDiscretePipes}\}}

\newcommand{\subHeatingSystem}{\text{HS}}
\newcommand{\Tshot}[1][]{\dTinf_{2\text{h}#1}}									
\newcommand{\Tscold}[1][]{\dTinf_{2\text{c}#1}}								
\newcommand{\Tphot}{\dTinf_{1}}							
\newcommand{\Tpcold}{\dTinf_{2}}							
\newcommand{\dTHeatingSystem}{\Tshot-\Tscold}							
\newcommand{\Qcons}{Q_{cons}}									
\newcommand{\qs}{\flow_{\subHeatingSystem}}										
\newcommand{\qp}{\gijEdge{\flow}}										
\newcommand{\Cmin}{C_{\text{min}}}										
\newcommand{\Cmax}{C_{\text{max}}}										
\newcommand{\Cstar}{C^*}										
\newcommand{\NTU}{NTU}											
\newcommand{\U}{U}												
\newcommand{\epsilonNTU}{\epsilon}								

\newcommand{\Ahexoriginal}{A}					

\newcommand{\nominal}{\mathrm{nom}}
\newcommand{\veCmin}[1][]{\ve{C}_{\text{min}#1}}
\newcommand{\veCmax}[1][]{\ve{C}_{\text{max}#1}}						
\newcommand{\veTshot}[1][]{\ve{\dTinf}_{2\text{h}#1}}									
\newcommand{\veTscold}[1][]{\ve{\dTinf}_{2\text{c}#1}}								
\newcommand{\veqs}[1][]{\ve{\flow}_{\subHeatingSystem#1}}										
\newcommand{\veheatHeatingSystem}[1]{\ve{{\heat}}_{\rad#1}}
\newcommand{\Tshotdesign}{\Tshot[,\nominal]}	
\newcommand{\veTshotdesign}{\veTshot[,\nominal]}	
\newcommand{\Tscolddesign}{\Tscold[,\nominal]}	
\newcommand{\veTscolddesign}{\veTscold[,\nominal]}	

\newcommand{\Tphotdesign}{\dTinf_{1\text{h},\nominal}}	
\newcommand{\veTphotdesign}{\ve{\dTinf}_{1\text{h},\nominal}}	
\newcommand{\Tpcolddesign}{\dTinf_{1\text{c},\nominal}}	
\newcommand{\veTpcolddesign}{\ve{\dTinf}_{1\text{c},\nominal}}	

In this section, the multi-period topology and design optimization problem is formulated, then the method for solving the topology optimization is described, and finally the time aggregation method used is outlined. 

\subsection{Notation}

DHNs are a network technology and can therefore be efficiently represented in a directed graph $\dirGraph(\setNodes,\setEdges)$, where $\setNodes$ is the set of all nodes and $\setEdges$ is the set of all edges in the graph. An overview of the graph notation used to formulate optimization problems in this paper can be found in the table \ref{tab:DHNnotation}.

\begin{table}[h!]
	\centering
	\caption{Graph notation used in this paper}
	\label{tab:DHNnotation}
	\begin{tabularx}{90mm}{lX}
		\toprule
		\textbf{Notation} & \textbf{Description} \\
		\midrule
		$\dirGraph = (\setNodes, \setEdges)$ & \gls{dhn} as a directed graph\\ 
		$\begin{aligned}[t]
			\setNodes = &\,\Npro \cup \Ncon\\ \cup\,&\Njun
		\end{aligned}$ & Set of nodes, including heat producers, consumers, and junctions.\\ 
		$\begin{aligned}[t]\setEdges = &\,\Epro \cup \Econ\\ \cup \,&\Epipe\end{aligned}$ & Set of edges, including heat producers, consumers, and pipes.\\
		$(\gi,\gj)$ or $\gi\gj$ & Directed edge going from node $\gi$ to node $\gj$ \\
		\bottomrule
	\end{tabularx}
\end{table}

Furthermore, the cardinality can be used to define the number of components $\n\in\mathbb{N}_0$ in each subset, e.g. the number of pipes in the network: $\npipes=\card{\Epipe}$. A compact definition of the following component numbers in the \gls{dhn} is given in table \ref{tab:setSizes}. 

\begin{table}[h!]
	\centering
	\caption{Sizes of the component sets in the \gls{dhn}}
	\label{tab:setSizes}
	\begin{tabularx}{\columnwidth}{lX}
		\toprule
		\textbf{Definition} & \textbf{Number of} \\
		\midrule	
		$\defnPeriods$ & Periods \\
		$\defnNodes$ & Nodes \\
		$\defnEdges$ & Edges \\
		 $\defnCon$ & Consumer\\ 
		 $\defnPro$ & Producer\\ 
		  $\defnpipes$ & Pipes\\ 
		
		\bottomrule
	\end{tabularx}
\end{table}

\subsection{DHN design as an optimization problem}

\newcommand{\defStateVar}{\ve{\stateVar}_{\timeVar} = \tp{\left[\tp{\ve{\flow}}_{\timeVar},\tp{\ve{\pressure}}_{\timeVar},\tp{\ve{\dTinf}}_{\timeVar}\right]}} 

When designing a new \gls{dhn}, the performance throughout its future operation must be considered during the design. Optimizing the topology and design therefore requires solving a multi-period optimization problem representing multiple operating points throughout the year. In the corresponding optimization problem, the project cost $\costFull \left(\designVar,\stateVar\right) \in\Real{}$ of the \gls{dhn} is minimized by choosing the placement of the pipes and their diameters $\defTopVar$ as well as the heat production capacity $\ve{\capVar}\in \Real{\nPro}$. To ensure a feasible operation, the operating variables of the heat consumer substations $\ve{\radValve}_{\timeVar}\in \Real{\nCon}$ and the heat producers $\ve{\prodInput}_{\timeVar} \in \Real{\nPro}$ in each period with $\timeVar\in \setPeriods$ have to be chosen. For ease of notation, the operational and design variables are combined into a vector $\defDesignVar$. The physical state of the network at each period $\timeVar$ is defined as $\defStateVar$, containing the volumetric flows $\defFlow$, node pressures $\defPressure$, and node and pipe outlet temperatures $\defTemp$.

Here $\setPeriods=\{1,\dots,\timeSlices\}$ is the set of time periods, where $\defnPeriods$ is the number of periods considered. For simplicity, the state of a quantity $\ve{a}\in\Real{ \n_{\mathrm{sg}}\cdot\nperiods}$ in a given period $\timeVar$ is denoted as $\ve{a}_\timeVar\in\Real{\n_{\mathrm{sg}}}$, where $\n_{\mathrm{sg}}\in\Real{}$ is the dimension of the quantity in a single period. The worst case period in which the network must ensure feasible operation is denoted as $\peakPeriod$. The multi-period topology and design optimization problem for \gls{dhn}s is thus

\begin{equation}
	\begin{aligned} \label{eq:optMP}
		\min_{\designVar,\stateVar} &\qquad
		\costFull \left(\designVar,\stateVar\right) &\\
		s.t.& \qquad \defModelConstraints = 0,\quad\forall \timeVar \in \setPeriods\,, &\\
		& \qquad \defStateConstraints \leq 0, & \\
		& \qquad \topVar\in\{0,\diameter\}^{\npipes}\,, & \\
		& \qquad \opVarBoxConstraints{\designVar}\,. &
	\end{aligned}
\end{equation}

where the network design variables $\designVar$ and the physical variables $\stateVar$ are optimized to satisfy the nonlinear model equations $\defModelConstraints$ representing the hydraulic and thermal transport problem in the network. Additional technological constraints (e.g. satisfying consumer heat demand or imposing maximum network pressure) are represented by $\defStateConstraints$. The binary topological choice of pipe placement is modeled by $\topVar\in\{0,\diameter\}^{\npipes}$, where the diameter of an existing pipe can be any value in the available continuous range. The design variables are bounded in $\designVarLower$ and $\designVarUpper$.

\newcommand{\npv}{NPV}
\newcommand{\npvConst}{f}
\newcommand{\npvN}{A}
\newcommand{\npvt}{k}
\newcommand{\npvDiscount}{e}

\newcommand{\npvConstCAPEX}{\npvConst_{\mathrm{\subCAPEX}}}
\newcommand{\npvConstOPEX}{\npvConst_{\mathrm{\subOPEX}}}

\newcommand{\npvCost}{C}

\newcommand{\actuarialRate}{\npvDiscount_{\mathrm{a}}}
\newcommand{\energyInflation}{\npvDiscount_{\mathrm{i}}}

\newcommand{\periodWeight}{\omega}

The discounted lifetime cost of the planned DHN $\costFull$ in this multi-period approach is defined as 
\begin{equation} \label{eq:totalCost}
	\begin{aligned} 
		\costFull\left(\designVar,\stateVar\right) &= \costi{\subPipe,\subCAPEX}\left(\designVar\right) + \costi{\subHeat,\subCAPEX}\left(\designVar\right) \\
		&+ \npvConstOPEX \sum_{\timeVar = 1}^{T} \periodWeight_{\timeVar} \left[\costi{\subHeat,\subOPEX}\left(\designVar,\stateVar_{\timeVar}\right) + \costi{\subPump,\subOPEX}\left(\stateVar_{\timeVar}\right)\right]  \,,
	\end{aligned}
\end{equation}

with $\npvConstOPEX = \sum_{\npvt=1}^{\npvN} \frac{1}{\left(1+\npvDiscount\right)^\npvt} $ assuming an investment horizon of e.g. $\npvN = 30~\mathrm{years}$ and a discount rate $\npvDiscount=5\%$. Here we consider the temporal weight $\ve{\periodWeight}\in\Real{\nperiods}$ of each period. We further assume that the chosen periods $\timeVar$ are sufficiently representative for a full year operation and $\sum_{\timeVar=1}^{\timeSlices}\periodWeight_{\timeVar} = 1$. The individual cost components of pipe \gls{capex} $\costi{\subPipe,\subCAPEX}$, heat production \gls{capex} $\costi{\subHeat,\subCAPEX}$, heat production \gls{opex} $\costi{\subHeat, \subOPEX}$, and pump \gls{opex} $ \costi{\subPump,\subOPEX}$ are described in previous work of the authors \cite{Blommaert2020a,Wack2022a}. Their complete definition as well as the complete optimization problem definition can be found in \ref{app:Cost}.

\subsection{Density-based topology optimization}
The challenging topology optimization problem of pipe placement $\topVar\in\{0,\diameter\}^{\npipes}$ is formulated as a density-based topology optimization problem. The investment cost of the pipes is defined as follows:
\begin{equation} \label{eq:Jpipe}
	\costi{\subPipe,\subCAPEX}\left(\topVar\right) = \sum_{\gi\gj \in \Epipe}\left(\Jpipepol_1  \gijEdge{\diameter} + \frac{1}{2}\bar{\Jpipepol}_0(\gijEdge{\diameter}) \right)\gijEdge{\length}\,, 
\end{equation}
with the interpolation coefficient $\Jpipepol_1\in\Real{}$ of a linear interpolation over the catalog cost per meter and the pipe length $\ve{\length}\in\Real{\npipes}$. In this density-based approach, the constant cost term of topological changes, $\bar{\Jpipepol}_0$, is formulated in a SINH-like manner, reducing the volumetric efficiency of small intermediate diameters: 

\begin{equation}
	\bar{\Jpipepol}_0(\gijEdge{\diameter})= \Jpipepol_0 \left(\frac{2}{(1+\exp(-\Jpipesmooth\left(\gijEdge{\diameter}-\diameterMin\right))})-1\right)\,,
\end{equation}

\begin{figure}[ht!]
	\centering
	\includegraphics{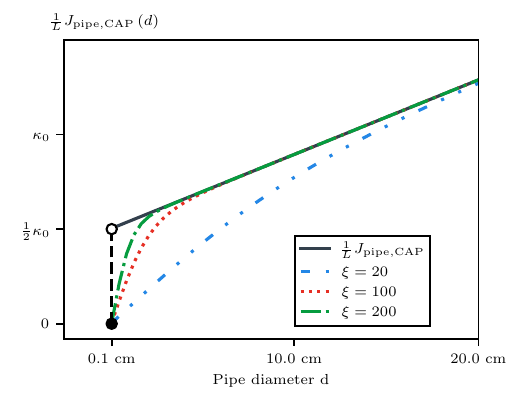}
	\caption{A SINH-like penalization approach for the fixed-term investment cost $\bar{\Jpipepol}_0$, which leads to a penalization of the investment cost $\cost_{\subPipe,\subCAPEX}$. Visualized for different values of the penalization parameter $\Jpipesmooth$.}
	\label{fig:penInvestmentCost2}
\end{figure}

$\defSetPipes$, with the penalty parameter $\Jpipesmooth$, the interpolation coefficient $\Jpipepol_0\in\Real{}$ and a minimum pipe diameter of $\diameterMin\in\Real{}$. This formulation is similar to that of Pizzolato et al. \cite{Pizzolato2018}. The penalized pipe cost function is visualized in figure \ref{fig:penInvestmentCost2}. This continuous density-based formulation avoids the direct solution of the challenging \gls{minlp} and allows the optimization of large heating networks while maintaining physical accuracy using nonlinear models. For a more detailed description of this density-based approach to \gls{dhn}s, the reader is referred to Wack et al. \cite{Wack2022a} and Blommaert et al. \cite{Blommaert2020a}.

\newcommand{\cHeatCAPEX}{\npvCost_{\mathrm{hC}}}
\newcommand{\cHeatCAPEXi}[1]{\npvCost_{\mathrm{hC},#1}}
\newcommand{\capacityFactor}{F}
\newcommand{\efficiency}{\eta}
\newcommand{\producerEfficiency}{\efficiency_{\pro}}

\newcommand{\maxCapacity}{P_{\textrm{max}}}
\newcommand{\prodCostVariable}{\npvCost_{\textrm{prod},1}}
\newcommand{\veprodCostVariable}{\\ve{npvCost}_{\textrm{prod},1}}
\newcommand{\prodCostFixed}{\npvCost_{\textrm{prod},2}}
\newcommand{\veprodCostFixed}{\ve{\npvCost}_{\textrm{prod},2}}
\newcommand{\prodCostOandM}{\npvCost_{\textrm{prod},3}}
\newcommand{\veprodCostOandM}{\ve{\npvCost}_{\textrm{prod},3}}

\subsection{A multi-period DHN optimization framework}
When minimizing the project cost $\costFull$ of a DHN by optimizing the DHN design and operation $\designVar$, the physics of future network operation must be accurately captured. Therefore, the optimization problem is constrained by a set of nonlinear model equations $\equalCon_{\timeVar}(\designVar_{\timeVar},\stateVar_{\timeVar})$ representing the hydraulic and thermal transport problem in the network. By assuming incompressible flow, DHNs can be hydraulically modelled as a sequence of steady states \cite{Stevanovic2007}. Thermal transients in the network, on the other hand, can last for several hours \cite{Benonysson1995}. The validity of the steady-state assumption therefore depends on whether each time period is long enough to allow thermal equilibrium to be reached for the majority of the time period. In this approach, periods are chosen that are representative of a full year, with the duration of each period being greater than the typical duration of thermal transients in DHNs, thus ensuring the validity of the steady state assumption. Furthermore, without considering thermal storage, a balance between supply and demand must be achieved in each period, allowing each time slice to be solved independently using a quasi-steady-state approach. The network modeled is based on the previous work of the authors \cite{Wack2022a,Blommaert2020a}, which models the conservation of mass, momentum, and energy for the heat suppliers, consumers and pipes in the network. The complete quasi steady-state model is described in \ref{app:model}. 

In the density-based approach, a nonlinear optimization problem is solved using an Augmented Lagrangian approach in combination with a Quasi-Newton method based on adjoint gradients. Similar to the forward problem, the adjoint problem of each period can be solved independently, which allows parallel computation of the optimization problem across time periods. The derivation of the independent computation of adjoint gradients for this multi-period framework can be found in \ref{app:adjoint}. For a comprehensive explanation of the density-based topology optimization approach and the nonlinear programming solver, the reader is referred to Wack et al. \cite{Wack2022a} and for the steady-state forward model and the adjoint gradient, to Blommaert et al. \cite{Blommaert2020a}.

\subsection{Temporal resolution and time aggregation}
Defining the topology and design optimization problem as time-dependent, using a multi-period approach, allows to consider the time dependence of a number of parameters that influence the design of \gls{dhn}s. The focus of this paper is on the temporal variations of consumer heat demand, the outdoor temperature, and the availability of intermittent heat sources, as they significantly affect the design of modern \gls{dhn}s.

\begin{figure*}[ht!]
	\centering
	\begin{subfigure}{90mm}
		\centering
		\includegraphics{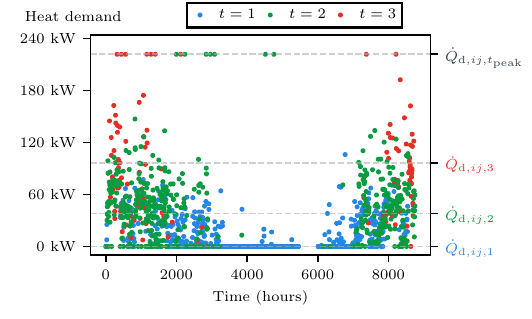}
		\caption{Commercial}
		\label{fig:timeSeries1a}
	\end{subfigure}
	\hfill
	\begin{subfigure}{90mm}
		\centering
		\includegraphics{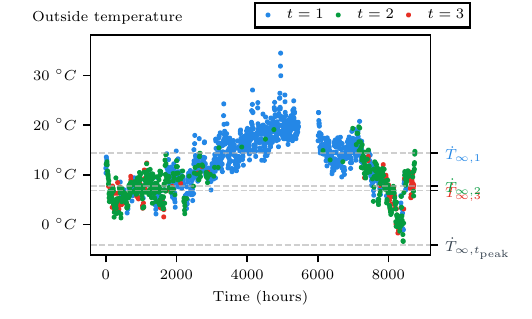} 
		\caption{Outdoor temperature}
		\label{fig:timeSeries1b}
	\end{subfigure}
	
	\caption{Illustration of time series aggregation for heat demand time series of a demand point in the a) commercial zone and b) the outdoor temperature. The heat demand time series are aggregated into three representative periods $\{\Qdemandi{,\gi\gj,1},\Qdemandi{,\gi\gj,2},\Qdemandi{,\gi\gj,3}\}$ and a peak period $\Qdemandi{,\gi\gj,\peakPeriod}$ using k-medoids clustering.}
	\label{fig:time_series_aggregation}
\end{figure*}

The time series of heat demand and other time-dependent parameters are often of high resolution. To manage the computational complexity, temporal aggregation methods are necessary to keep the problem tractable while capturing the essential features of the time-dependent problem. As discussed earlier, there are a variety of methods commonly used in \gls{dhn} modeling and design. In this study, the k-medoids clustering approach described by Kotzur et al. \cite{Kotzur2018} is used to aggregate the heat demand and outdoor temperature profiles into a number of representative periods. Figure \ref{fig:timeSeries1a} and figure \ref{fig:timeSeries1b} provide a visualization of the time series data and the resulting aggregated periods.

\newcommand{\nAttributes}{\n_{\mathrm{at}}}
\newcommand{\nObservations}{\n_{\mathrm{ob}}}
\newcommand{\nCluster}{\n_{\mathrm{cl}}}
\newcommand{\veQdemand}[1][]{\ve{\heat}_{\mathrm{d}#1}}
Before aggregating the time series, they are normalized by attribute type, such as heat demand or outdoor temperature, following Kotzur et al. \cite{Kotzur2018} so that all time series can be evaluated on the same scale. In this paper, the Matlab toolbox \textit{kmedoids} \cite{matlabkmedoidsBibTex} is used to perform the clustering.

Aggregating the time series has the disadvantage of potentially truncating the so-called peak period, since it is not representative of a group or cluster of periods \cite{Kotzur2018}. To ensure that the optimized network design remains feasible under worst-case conditions, an additional worst-case period $\peakPeriod\in\setPeriods$ is added to the set of aggregated periods. Here, the worst case of each attribute, e.g. the peak demand of all consumers and the lowest outdoor temperature, is combined and added as an additional period to the optimization. The duration and the period weight $\periodWeight$ of this peak period are difficult to quantify. Therefore, only the \gls{capex} cost and feasibility of this peak period is considered by setting the period weight $\periodWeight_{\peakPeriod}=0$. This ensures that the optimized \gls{dhn} design can meet the heat demand of all consumers even under worst case conditions.

\section{Density-based topology optimization of a heating network based on multiple periods - a case study}\label{chap:simple}

\begin{figure*}[ht!]
	\centering
	\includegraphics{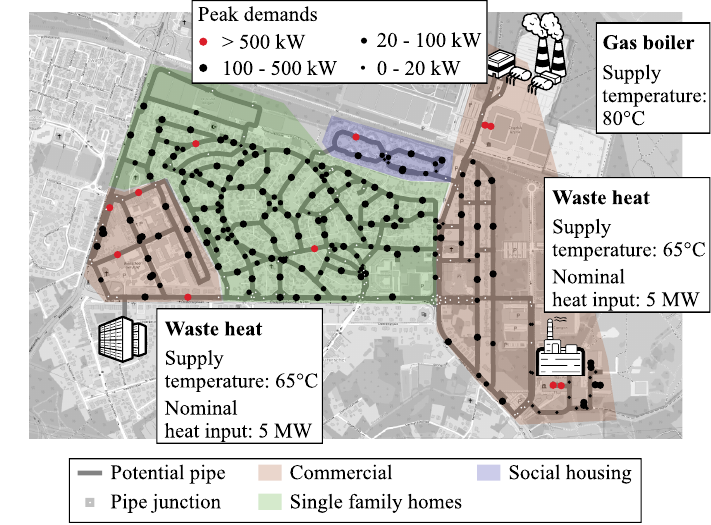}
	\caption{View of the Waterschei neighborhood in Genk, Belgium, showing the location of the waste heat sources (southeast and southwest) and the proposed peak gas boiler (north), along with the distribution of the 217 aggregated demand points.}
	\label{fig:neighborhood}
\end{figure*}

We now study what an automated design approach based on a multi-period-based topology optimization framework can do for the design process of \gls{dhn}s. For this purpose we consider a \gls{dhn} development project in the Waterschei neighborhood of Genk, Belgium. A visualization of the case study setup is provided in figure \ref{fig:neighborhood}. In this case study for a mostly residential neighborhood consisting of about 3800 buildings, a heating network is planned that potentially connects the buildings to two fictitious waste heat sources, an old mine shaft in the southeast, and waste heat from a commercial site in the southwest. To provide the remaining heat, a peak gas boiler is planned to be built in the north of the neighborhood. 

\definecolor{redNice}{HTML}{D88B84}
\definecolor{blueNice}{HTML}{8B84D8}
\definecolor{greenNice}{HTML}{84D88B}

\subsection*{Case setup}

For this case study, the existing street grid and heat demand locations are used as input to the optimization. Both the heat demand location and the street grid of the neighborhood were derived from \gls{gis} data (as described in Salenbien et al. \cite{Salenbien2022}). The total annual heat demand of the houses in the neighborhood was derived from the \textit{Warmtekaart Vlaanderen} database \cite{warmtekaart} and spatially aggregated per street segment into 217 representative demands \cite{Salenbien2022}. To better represent the different characteristics of the dwellings in the neighborhood, they were divided into a commercial zone (figure \ref{fig:neighborhood} red area), single family houses (green area), and social housing (blue area). For each zone, representative annual heat demand curves were generated using the IDEAS library \cite{Jorissen2018} and scaled to the individual peak demand of each building. The occupancy behavior of these buildings was modeled using the StROBe library \cite{Baetens2016}. An example of these time series for the commercial zone is visualized in figure \ref{fig:timeSeries1a}. For the time-dependent outside temperature $\ve{\TOutside}\in\Real{\nperiods}$ an hourly time series of temperatures for the year 2022 has been obtained from the Royal Meteorological Institute of Belgium for a weather station in Flanders \cite{RMImeteo2023BibTex}. This time series is shown in figure \ref{fig:timeSeries1b}.

As production sources, two fictitious waste heat sources located to the southeast and southwest of the neighborhood are considered, providing $\SI{5}{\mega\watt}$\footnote{The design heat supply is referenced to a design return temperature of $\SI{20}{\degreeCelsius}$} of heat at $\SI{65}{\degreeCelsius}$. To provide the remaining heat, a peak gas boiler should be sized and built in the north of the neighborhood to provide heat at $\SI{80}{\degreeCelsius}$. To design the topology of the heating network, the street grid of the neighborhood is considered as potential routes for the pipes. A list of the main assumptions and parameters of the optimization problem can be found in table \ref{tab:properties}.

\begin{table*}[h!]
	\centering
	\caption{Key parameters of the optimization problem used for the multi-period case study.}	
	\label{tab:properties}
	\begin{tabularx}{\textwidth}{l|ll|X}		
		\textbf{Property} & \textbf{Value} & \textbf{Unit} & \textbf{Remarks}\\
		\hline\hline
		$\npvN$ & 30 & years &\\
		$\interestRate$ & 0.05 & - &\\
		$\TOutside \in$ &  $\{14.37,7.80,...$& \si{\degreeCelsius}&Belgian weather station\\	
		&$6.85,-4.11\}$&&2022 \cite{RMImeteo2023BibTex}\\	
		$\density$ &  $983$&\si{\kilogram\per\meter^3}&\\
		$\viscosity$ &  $4.67\times10^{-4}$&\unit{\pascal\second}&\\
		$\spHeatCap$ &  4185 &\si{\joule\per{\kilogram\kelvin}}&\\
		$\condGround$ & 1&\si{\watt\per{\meter\kelvin}}&Wet ground \cite{solidsConductivityBibTex}\\
		$\condInsul$ &  0.0225&\si{\watt\per{\meter\kelvin}}&Hard foam insulation \cite{isoplusCatalouge}\\
		$\depthPipe$ &  1&\si{\meter}&\\
		$\ratioInsul$ &  1.87&-&Average catalogued ratio \cite{isoplusCatalouge}\\
		\hline
		$\Jpipepol_0$ & 501.3&\si{\sieuro\per\meter}&\\
		$\Jpipepol_1$ & 1976.3&\si{\sieuro\per\meter\squared}&\\
		$\cPumpOPEX$ & 0.1&\si{\sieuro\per{\kilo\watt\hour}}&\\
		$\pumpEff\in$ & $\{0.81,0.81\}$ &&\\
		$(\dTinf_{\mathrm{ref}})_{\gi\gj} \in$ &  $\{20,20\}$&  $\unit{\degreeCelsius}$&\\
		$\maxPressure$ & 1&\si{\mega\pascal}&\\			
		\hline
		$\THouse$ & 20&\si{\degreeCelsius}&\\
		$\Tphotdesign,\Tpcolddesign$ & 60,42 & \si{\degreeCelsius}&\\
		$\Tshotdesign,\Tscolddesign$ & 55,40&\si{\degreeCelsius}&\\
		\hline
		Gas boiler &&&\\
		$\prodCostVariable $ &  $225$  &\si{\sieuro\per{\kilo\watt}}&Heinen et al. \cite{Heinen2016}\\
		$\prodCostFixed$ &  $2200$ &\si{\sieuro}&Heinen et al. \cite{Heinen2016}\\
		$\prodCostOandM$ & 
		$235$ &\si{\sieuro\per{\year}}&Heinen et al. \cite{Heinen2016}\\
		$\cHeatOPEX$ &  $0.0319$  &\si{\sieuro\per{\kilo\watt\hour}}&Average natural gas price in EU for non-household consumers from 2019-2022 \cite{eurostatBibTex}\\
		\hline
		Waste heat&&&\\
		$\prodCostVariable $ \& $\prodCostFixed$ & 0&\si{\sieuro\per{\kilo\watt}}\,\&\,\si{\sieuro}&\\
		$\prodCostOandM$ & 0&\si{\sieuro\per{\year}}&\\
	\end{tabularx}
\end{table*}

\newcommand{\conversionFactor}{K}

After preprocessing and aggregation, the optimization problem has 217 heat consumers, 847 rout segments for potential piping, $\timeSlices=4$ time periods, and a total of 2384 design and operational variables. The time series were pre-processed to remove a large consecutive time span of inactivity without demand in the summer of 23.0 days, resulting in $\conversionFactor=8208.5 \si{\hour\per\year}$ of active operating hours throughout the year. The time series were then aggregated into $\timeSlices=4$ representative time periods with period weights $\ve{\periodWeight}=\{0.650,0.270,0.080,0\}$, each corresponding to $\{222.3, 92.3,27.4,0\}$ days of a representative year. A study has been performed to validate the independence of the design from the chosen number of periods and can be found in the \ref{app:validation}. The design optimization problem for this case study is solved in the next section using the automated design approach. 

\clearpage
\subsection{Worst-case versus multi-period design}
\begin{figure}[ht!]
	\centering
	\includegraphics[width=90mm]{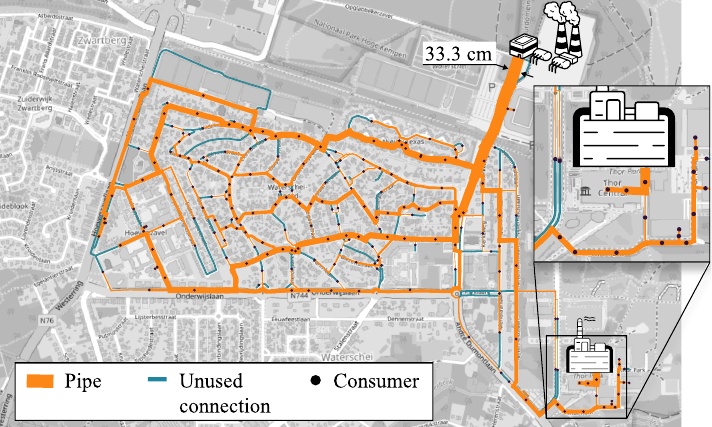}
	\caption{Optimized network topology and pipe sizing for a \gls{dhn} considering only worst case conditions. The line thickness represents the installed pipe diameter. The waste heat source in the southeast supplies only two heat consumers in the immediate vicinity.}
	\label{fig:peak_topology}
\end{figure}

First, the \gls{dhn} is optimized considering only a worst case scenario where $\timeVar = \peakPeriod$, similar to what was done previously in Wack et al. \cite{Wack2022a}. The topology optimization approach is used to optimize the network topology, pipe sizes, and required peak boiler capacity. In this first study, a single waste heat source located in the southeast is considered for the network along with the gas boiler located in the north. The optimal network topology for peak conditions is visualized in figure \ref{fig:peak_topology}. When only the peak conditions are considered in the optimization, the network is designed to satisfy the maximum load at the lowest cost, resulting in two separate branched networks for the gas boiler and the waste heat source. 

In this study, this results in the waste heat source serving only 3 nearby consumers. These 3 commercial consumers have a combined peak demand of $\Qdemand= \SI{2.751}{\mega\watt}$ and consume the available heat of $\heat_{\pro}= \SI{2.717}{\mega\watt}$ from the waste heat source at a return temperature of $\temperature=\SI{40.5}{\degreeCelsius}$. In fact, this pure worst-case analysis would suggest using the waste heat source locally in the immediate neighborhood as an `energy island' rather than integrating it into the larger neighborhood distribution network.

Now, evaluating the physics and cost for this optimized worst-case design, considering all periods $\timeVar \in \setPeriods$, highlights the increase in temporal detail added with a multi-period approach. Considering $\timeSlices=4$ periods accurately approximates the annual energy use of the neighborhood of $\SI{3.67}{\giga\watt\hour\per\year}$ with only a $\SI{1.09}{\percent}$ error. This increased accuracy when considering multiple periods results in a more accurate estimate of the network operating cost of heat of $\SI{19.9}{\mega\sieuro}$ compared to $\SI{185.3}{\mega\sieuro}$ when considering only a single worst case. While multi-period network evaluation provides improvements in modeling accuracy over worst-case analysis, the more significant impact of the multi-period approach is the added spatial and temporal detail for topology and design optimization. This impact is further explored in the following section.

\clearpage
\subsection{Impact of the automated multi-period approach on the DHN design}\label{sec:MPworstvsMulti}

\begin{figure}[ht!]
	\centering
	\includegraphics[width=90mm]{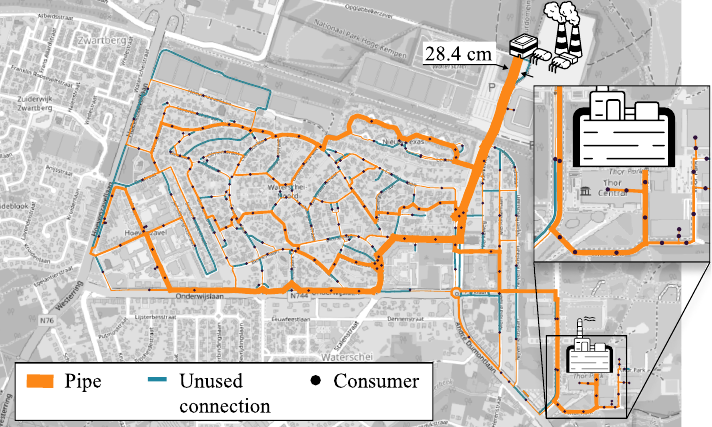}
	\caption{Optimized network topology and pipe sizing for a \gls{dhn} considering multiple periods. The line thickness represents the installed pipe diameter. A backbone connects the waste heat source and the gas boiler, allowing for greater integration of waste heat into the network.}
	\label{fig:3period_topology}
\end{figure}

Now the network topology, pipe sizes, and required peak boiler capacity are optimized as a full multi-period optimization with $\timeVar \in \setPeriods$, considering three representative periods and worst case feasibility ($\timeSlices=4$). The optimal topology of the \gls{dhn} is shown in \ref{fig:3period_topology}. Unlike the previous worst-case design (figure \ref{fig:peak_topology}), considering multiple periods in the optimization results in a single integrated network topology that is simultaneously designed for different operating conditions throughout the year. The multi-period approach allows for a more efficient network design while ensuring adequate capacity during peak demand periods. In this multi-period optimization, the network topology is now built around a backbone connecting the peak boiler and waste heat source. This allows for spatial shifting of loads between the two sources under different operating conditions. 

\begin{figure}[ht!]
	\centering
	\includegraphics[width=90mm]{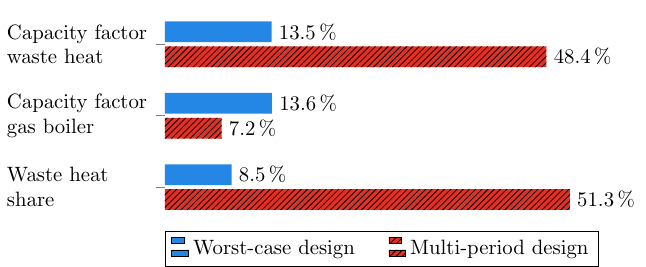}
	\caption{Comparison of production plant utilization for worst-case and multi-period design. The multi-period optimization increases the networks waste heat share and decreases the necessary operation time of the gas boiler}
	\label{fig:peakVsMPDifferenceProducer}
\end{figure}

This single integrated network topology allows an overall higher share of the waste heat source to be integrated, especially during low demand periods. The waste heat share for both the worst-case and multi-period designs is compared in figure \ref{fig:peakVsMPDifferenceProducer}. To ensure a fair performance comparison of both network designs, the performance of the optimized worst-case design is also simulated for all representative periods $\timeVar \in \setPeriods$. The integrated network topology resulting from the multi-period optimization leads to an overall increase in the waste heat share of the network from $\SI{8.5}{\percent}$ for the worst-case design to $\SI{51.3}{\percent}$ for the multi-period design. This shift towards integrated annual operation further increases the capacity factor of the waste heat source and consequently decreases the capacity factor of the gas boiler. 

\begin{figure}[ht!]
	\centering
	\includegraphics[width=90mm]{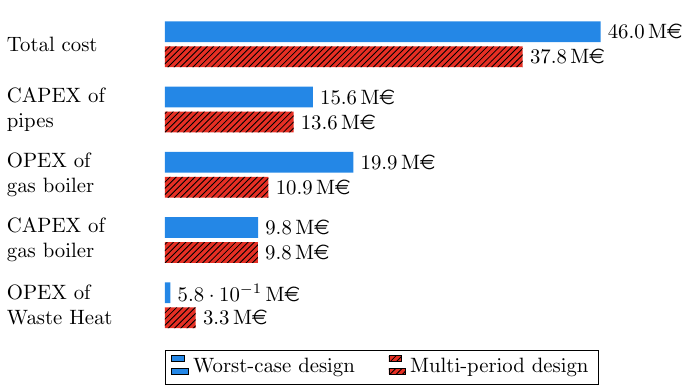}	
	\caption{Cost comparison of worst-case and multi-period design. The increased waste heat share of the multi-period design leads to a drop in the operational cost of the gas boiler and to a reduction in piping costs.}
	\label{fig:peakVsMPDifference}
\end{figure}

Since the \gls{opex} of the waste heat source is assumed to be cheaper than the peak boiler, this shift results in a decrease of the peak boiler \gls{opex} by \SI{8.95}{\mega\sieuro}, while the waste heat \gls{opex} increases by \SI{2.68}{\mega\sieuro}. The cost changes from worst-case to multi-period design are visualized for the different cost components in figure \ref{fig:peakVsMPDifference}. This large reduction in heat \gls{opex} cost in the multi-period design is the largest contributor to the overall reduction in total project cost of $\SI{8.22}{\mega\sieuro}$ or $\SI{17.9}{\percent}$ compared to the worst-case design. A constant peak boiler \gls{capex} of \SI{9.84}{\mega\sieuro} is observed for both designs to ensure operational feasibility in both worst-case and multi-period designs. This shows that the proposed automated multi-period design approach results in integrated, cost-effective networks that are also conservatively sized to be able to operate under worst-case conditions.

\begin{figure*}[ht!]
	\centering
	\begin{minipage}[b]{90mm}
		\centering
		\includegraphics[width=90mm]{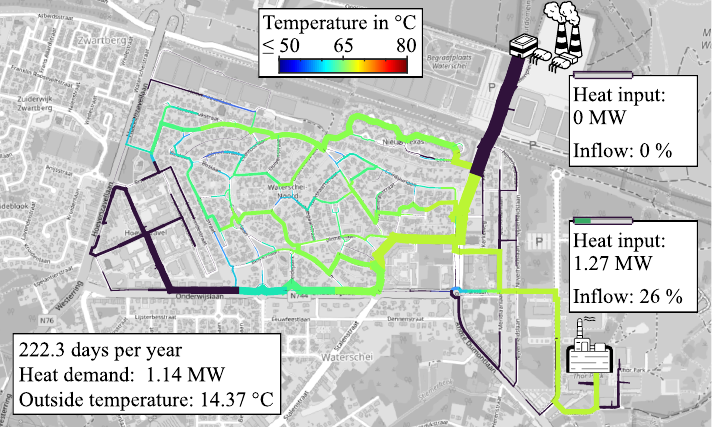}
		\subcaption{Period 1}
		\label{fig:temp_distribution_period1}
	\end{minipage}\hfill
	\begin{minipage}[b]{90mm}
		\centering
		\includegraphics[width=90mm]{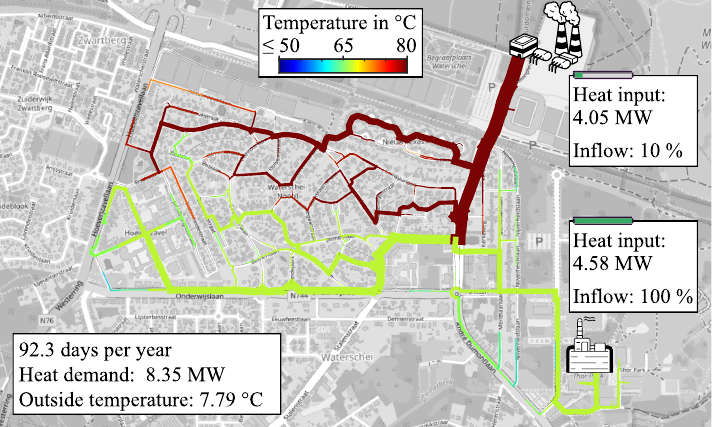}
		\subcaption{Period 2}
		\label{fig:temp_distribution_period2}
	\end{minipage}
	
	\vspace*{\floatsep}
	\caption{Temperature distribution of the optimized \gls{dhn} for for period 1 and 2. The line thickness represents the pipe sizes, while the color represents the water temperature of the pipes.}
\label{fig:temperature_distributions_comparison_a}
	\end{figure*}
	
\begin{figure*}[ht!]
	\centering	
	\begin{minipage}[b]{90mm}
		\centering
		\includegraphics[width=90mm]{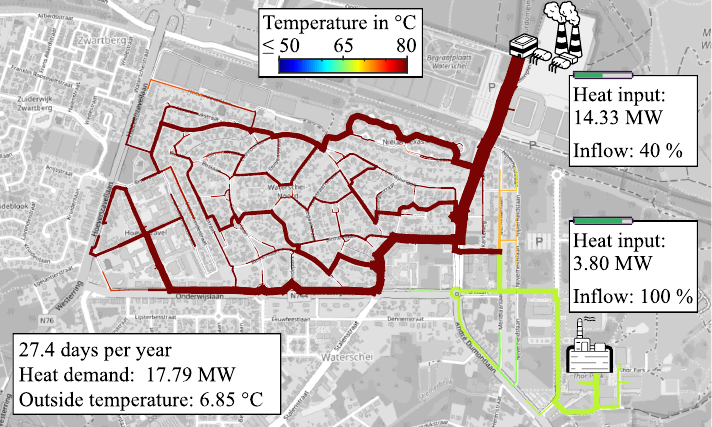}
		\subcaption{Period 3}
		\label{fig:temp_distribution_period3}
	\end{minipage}\hfill
	\begin{minipage}[b]{90mm}
		\centering
		\includegraphics[width=90mm]{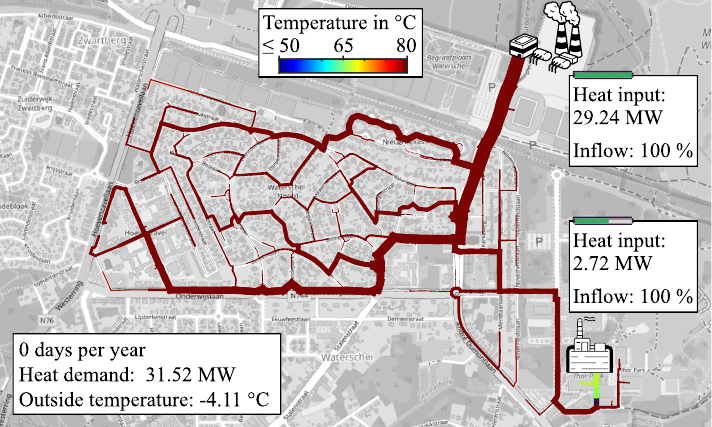}
		\subcaption{Peak period $\peakPeriod$}
		\label{fig:temp_distribution_peak_period}
	\end{minipage}
	\caption{Temperature distribution of the optimized \gls{dhn} for period 3 and the peak period. The line thickness represents the pipe sizes, while the color represents the water temperature of the pipes.}
	\label{fig:temperature_distributions_comparison_b}
\end{figure*}

The spatial load shift from the peak boiler to the waste heat source in the multi-period design can be further observed by analyzing the network temperatures for the multi-period design over the different periods in figure \ref{fig:temp_distribution_period1}, \ref{fig:temp_distribution_period2}, \ref{fig:temp_distribution_period3}, and \ref{fig:temp_distribution_peak_period}. The water temperature is a good indicator of which production site is supplying a consumer. During the peak demand period visualized in figure \ref{fig:temp_distribution_peak_period}, the network experiences the highest heat demand. During this period, the waste heat source serves only the demand in its immediate vicinity, and the peak gas boiler plays a crucial role by providing the remaining heat needed in the network. During periods of lower demand, as visualized in figure
\ref{fig:temp_distribution_period1}, \ref{fig:temp_distribution_period2}, and \ref{fig:temp_distribution_period3}, the cheaper low-temperature waste heat source in the southeast of the neighborhood is able to meet an increased portion of the heat demand. 

The visualization of the water temperatures of period 3 in figure \ref{fig:temp_distribution_period3} also shows that during off-peak periods, water from the waste heat source is mixed with water from the gas boiler in the westernmost branch of the network (see the orange branch). This effectively boosts the temperature of the water downstream and maximizes the use of waste heat during this period. The supply from the waste heat source is not sufficient to meet all the demand after the mixing point, so the mixed-in gas boiler heat can supply the remainder. While this mixing operation may be challenging to control in practice, this result could motivate the integration of valves or substations at these mixing points to facilitate the operation of networks aimed at maximizing the waste heat share.  

The cost comparison in figure \ref{fig:peakVsMPDifference} also shows a cost reduction of \SI{2.04}{\mega\sieuro} for the \gls{capex} of the pipe infrastructure. This reduction is achieved because the optimized multi-period design installs smaller pipe diameters in the network overall than the optimized worst-case design. While the worst-case design installs an average pipe diameter of \SI{9.37}{\centi\meter}, the multi-period design is able to install an average pipe diameter of \SI{7.29}{\centi\meter} in the network\footnote{The average network diameter was calculated for each pipe installed in the optimized network, weighted by pipe length.}. In the worst-case design, the optimization cannot take advantage of the different load scenarios to maximize waste heat utilization and reduce costs. The optimal topology and especially the pipe diameters of the network are therefore designed to meet most of the heat demand with the gas boiler. The multi-period design, on the other hand, operates the network at a lower overall load due to a more accurate approximation of the annual energy consumption and operates at lower temperatures. This new balance between pump, heat and piping costs results in overall smaller diameters for the multi-period design. 

\clearpage
\subsection{Accounting for waste heat unavailability} \label{sec:intermittancy}

\begin{figure}[ht!]
	\centering
	\includegraphics[width=90mm]{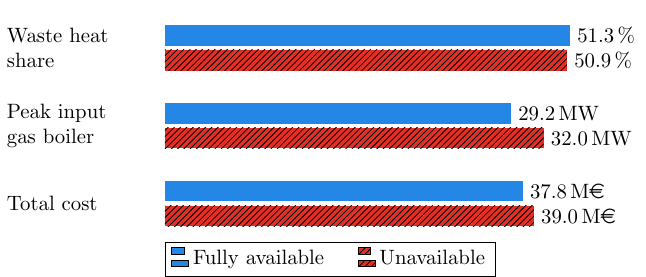}
	
	\caption{Comparison of key performance metrics of the optimized \gls{dhn} design considering the unavailability of waste heat.}
	\label{fig:intermitancy_comparison}
\end{figure}

While the previous study focused on the temporal variations of heat demand and outdoor temperature, the multi-period formulation also allows to prevent the unavailability of heat suppliers from affecting the ability to deliver heat throughout the network. To illustrate this, a study is performed where the waste heat source is considered unavailable during the peak period $\peakPeriod$. The network topology and design is then optimized again taking this unavailability into account. An optimized network design is found that is still able to deliver heat when the waste heat sources are unavailable. Its key performance metrics are visualized in Figure \ref{fig:intermitancy_comparison}. The results show that the peak boiler input in the peak period is increased from $\SI{29.2}{\mega\watt}$ to $\SI{32.0}{\mega\watt}$ to counteract the unavailability of the waste heat sources. This increase in peak boiler capacity leads to an increase in the lifetime cost of the project by $\SI{3.1}{\percent}$ to $\SI{39}{\mega\sieuro}$. Despite the redundant design, a high waste heat share of $\SI{59.5}{\percent}$ can be maintained. This shows how resolving the intermittency of waste heat availability in the optimization can create redundancy in the network design without sacrificing a high waste heat share.  

\clearpage
\subsection{Increasing connectivity of DHN topologies as spatial complexity increases}

The transition of the automated design approach based on density-based topology optimization from a worst-case optimal design to a multi-period approach also shows a progression from separate branched networks for gas boiler and waste heat source for the worst-case design (see figure \ref{fig:peak_topology}) to a single integrated network topology connecting both producers through a backbone for the multi-period design (see figure \ref{fig:3period_topology}). It is important to emphasize here that the optimization problem formulation of the proposed automated design approach does not impose any specific requirements on the network topology (e.g., branched or meshed network topologies), but the optimization problem allows these structures to emerge freely within the boundaries of the road network, based on the economic needs of the network project.

\begin{figure}[ht!]
	\centering
	\includegraphics[width=90mm]{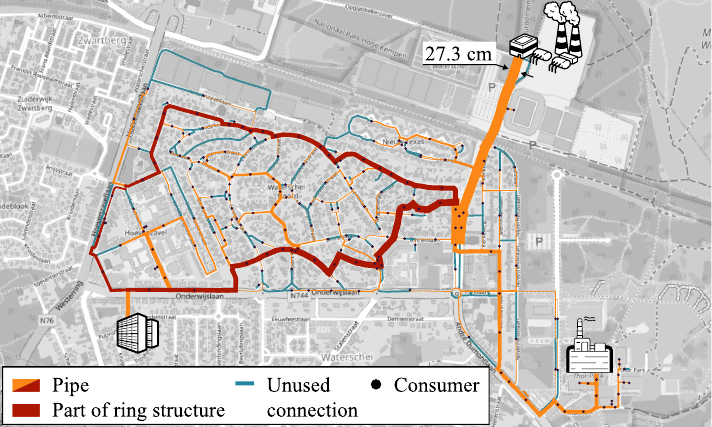} 
	\caption{Optimized network topology and pipe sizing for a \gls{dhn} with two waste heat sources. The line thickness represents the installed pipe diameter. In the optimized design, a ring structure colored red connects the southwestern waste heat source with the other producers. This structure was not imposed a priori.}
	\label{fig:topology_twoWasteHeatSources}
\end{figure}

Increasing the number of heat producers by adding a second waste heat source in the southeast of the neighborhood, which again provides a maximum of $\SI{5}{\mega\watt}$ at $\SI{65}{\degreeCelsius}$, further increases the connectivity of the network topology. The optimized network topology considering three heat producers in the multi-period design is visualized in figure \ref{fig:topology_twoWasteHeatSources}. This topology shows the emergence of meshed structures in the optimal network topology. Here we see a ring structure highlighted in red that connects the western waste heat source to the central backbone. This increased network connectivity allows for more flexible shifting of heat loads between different producers and heat consumers, resulting in more cost effective use of the heat supply. Such ring structures are part of the discussion on appropriate topologies for \gls{4gdh} networks with multiple heat sources. In particular, they are said to alleviate the capacity limitations of traditional tree structures \cite{Lund2018}. Their appearance as optimized topologies in an automated design tool such as the one proposed in this work may contribute to the discussion of future \gls{dhn} topologies.

The integration of a second waste heat source shows a further waste heat share increase of $\SI{30.4}{\percent}$ to $\SI{81.7}{\percent}$ and a subsequent further decrease of the total project cost by $\SI{15.4}{\percent}$ to $\SI{31.987}{\mega\sieuro}$. 

The resulting optimized network designs of this case study connect multiple (low) temperature heat sources at different temperatures in a single integrated meshed network design with high waste heat utilization and redundancy in heat production. It highlights the impact that the proposed automated multi-period design approach can have on the design process of modern next-generation \gls{dhn}s. It combines a density-based topology optimization approach, which allows for high physical accuracy while remaining scalable, with a multi-period approach, which allows the optimization to directly exploit temporal detail in operation. This ultimately enables the automated, cost-effective design of next-generation \gls{dhn}s with complex supply and demand characteristics by integrating them into meshed network topologies.

\clearpage
\section{Conclusions and outlook}
This paper proposes an optimization approach for the optimal topology and design of \gls{dhn}s that combines density-based topology optimization for \gls{dhn}s with a multi-period approach, allowing temporal variations in operation and key parameters to be directly integrated into the design process. The computational complexity of the time resolution is handled by a k-medoids clustering approach, that reduces the transient heat transfer problem to a limited number of steady-state simulations. 

A case study showed a clear difference between the optimized \gls{dhn} design based on a worst case and a multi-period design. The worst-case design proposed a purely local use of waste heat in the immediate vicinity of the waste heat source and resulted in a separate branched network for each heat producer. The multi-period approach, in contrast, designed a more efficient, integrated network that is simultaneously optimized for different operating conditions throughout the year, while ensuring adequate capacity during peak demand periods. This led to a reduction in the total project cost of $\SI{8.22}{\mega\sieuro}$ or $\SI{17.9}{\percent}$. Further consideration of producer unavailability in the optimal design showed that the approach can be used to automatically design redundant, cost-effective \gls{dhn}s that can handle heat source unavailability without sacrificing a high waste heat share of $\SI{59.5}{\percent}$.

The transition of the automated design approach from a worst-case to a multi-period approach showed a network design progression from separate branched networks in the worst-case design to a single integrated network topology connecting all producers for the multi-period design. The connectivity of the single integrated network continues to increase as additional heat producers are added. This increases the connectivity of the network and allows for more flexible shifting of heat loads between different producers and heat consumers, resulting in an overall more cost-effective use of the heat supply. The emergence of integrated and meshed topologies in this automated design approach, without imposing such structures a priori, indicates the importance that meshed network topologies may have for cost-effective future \gls{dhn}s.

The automated design tool presented here combines a density-based topology optimization approach, which allows for high physical accuracy while remaining scalable, with a multi-period approach. It is an important prerequisite for optimizing \gls{dhn}s that integrate intermittent renewable energy sources. Such design tools may ultimately enable the automated and cost-effective design of next-generation \gls{dhn}s that rely on matching variable heat demands to a combination of heat sources with intermittent characteristics in meshed network topologies that provide redundancy.

Future research should further investigate the effect of time aggregation on the optimal design of \gls{dhn}s and attempt to gain a more complete understanding of the required number of representative time periods. To increase the applicability of automated design for \gls{dhn}s, other potential design factors of \gls{dhn}s, such as different types of heat producers or the robustness of the optimized design should be considered. This will ultimately allow the automated design of renewable-based \gls{4gdh} networks.

\section{Data Availability}
A data-set including the structure, input parameters, time series and optimization results of the heating networks used in the case studies of this paper is available at the following link: \url{https://doi.org/10.48804/4OQJGR}. The optimization results can be replicated using the methodology and formulations described in this paper.
\section{Acknowledgements}
Yannick Wack is funded by the flemish institute for technological research
(VITO). 

Martin Sollich has received funding from the KU Leuven with the reference STG/21/016.

The authors would like to extend their gratitude to Anouk Robbeets for her input to this work. 
\section{CRediT authorship contribution statement}
\textbf{Yannick Wack}: Conceptualization, Methodology, Software, Formal analysis, Visualization, Writing – original draft

\textbf{Martin Sollich}: Investigation, Methodology, Software, Formal analysis, Writing – review \& editing

\textbf{Robbe Salenbien}: Conceptualization, Data Curation, Funding acquisition, Writing – review \& editing

\textbf{Martine Baelmans}: Conceptualization, Funding acquisition, Writing – review \& editing

\textbf{Jan Diriken}: Data Curation, Writing – review \& editing

\textbf{Maarten Blommaert}: Conceptualization, Methodology, Software, Supervision, Funding acquisition , Writing – review \& editing 

\appendix
\section{Detailed definition of the topology optimization problem for DHNs}\label{app:model}

To ensure consistency and reproducibility, a brief summary of the full topology optimization problem for DHNs is provided in this section. For a more detailed definition, interested readers are referred to Wack et al. \cite{Wack2022a} and Blommaert et al. \cite{Blommaert2020a}.

\subsection{Cost Function}\label{app:Cost}

\newcommand{\dTProducer}{\Delta\dTinf} 
The time invariant cost of the production capacity in the time dependent optimization problem is defined as:

\begin{equation}\label{eq:HeatCAPEX}
	\costi{\subHeat,\subCAPEX}\left(\designVar,\stateVar\right) = \sum_{\gi\gj \in \Epro}\gijEdge{\capVar} \gijEdge{\left(\maxCapacity \prodCostVariable + \prodCostFixed\right)} .
\end{equation}

Here $\ve{\capVar}$ is the heat production capacity variable and $\ve{\maxCapacity}\in\Real{\nPro}$ is the maximum capacity of each producer. $\veprodCostVariable\in\Real{\nPro}$ is the investment cost per unit of heat production capacity in $\si{\sieuro\per{\kilo\watt}}$, and $\veprodCostFixed\in\Real{\nPro}$ is the fixed investment cost in $\si{\sieuro}$.  The operational heat cost of the peak gas boiler is calculated using
\begin{equation}\label{eq:HeatOPEX}
	\begin{aligned}
		\costi[,\timeVar]{\subHeat,\subOPEX}\left(\designVar,\stateVar_{\timeVar}\right) =  \density \spHeatCap\conversionFactor\sum_{\gi\gj \in \Epro}&\gijEdge{\bigg(\cHeatOPEX \,\flow_{\timeVar} \, \dTProducer_{\timeVar} \\ &+ \capVar\,\prodCostOandM\bigg)}\, ,
	\end{aligned}
\end{equation}
with the unit price of heat $\ve{\cHeatOPEX}\in\Real{\nPro}$ in $\si{\sieuro\per{\kilo\watt\hour}}$, the conversion factor $\conversionFactor\in\Real{}$ in $\si{\hour\per\year}$, which defines the networks number of active operating hours during the year. The maintenance and fixed operating costs are given by $\veprodCostOandM\in\Real{\nPro}$ in \euro/year.  An additional state constraint is defined to ensure that the heat capacity of the producer $\ve{\capVar}$ is always greater than the required heat input per period. This constraint is defined as:

\begin{align}\label{eq:capacityConstraint}
	&-\left(\gijEdge{\capVar} - \gijtEdge{\dTProducer}\gijEdge{\left(\frac{ \flow_\timeVar  \producerEfficiency}{\maxCapacity}\right)} \spHeatCap \rho \right) \leq 0\,\\
	&\forall\gi\gj \in \Epro \,.
\end{align}

The operational cost of pumps at the heat production sites is computed with
\begin{equation}\label{eq:pumpOPEX}
	\costi[,\timeVar]{\subPump,\subOPEX}\left(\stateVar_{\timeVar}\right) = \frac{1}{\pumpEff} \,\conversionFactor\sum_{\gi\gj \in \Epro}\cPumpOPEX\left(\gjtNode{\pressure}-\pressure_{\gi,\timeVar}\right) \gijtEdge{\flow} \, ,
\end{equation}
where $\pressure_{\gj,\timeVar}$ with $\gj\in \NproR$ represents the corresponding pressure at the feed node of a producer. The unit pumping price is defined by the electricity price $\cPumpOPEX\in\Real{}$ in $\si{\sieuro\per{\kilo\watt\hour}}$ and the pump efficiency is given by $\pumpEff\in\Real{}$. 

The optimization problem is constrained by a set of nonlinear model equations $\defModelConstraints=0$, representing the hydraulic and thermal transport problem in the network. It describes the conservation of mass, momentum, and thermal energy in the components of a heating network. The individual models are described in the following sections.
\subsection{Pipe model}
To model the momentum equations over a pipe, the empirical Darcy-Weisbach equation is used, which models the viscous pressure drop in incompressible flow as a function of the volumetric flow rates $\gijtEdge{\flow}$ through pipes of length $\ve{\length}$ \cite[p.120]{strömungslehre}: 

\begin{equation} \label{eq:pipeMomentum}
	(\gitNode{\pressure} - \gjtNode{\pressure}) =  \gijtEdge{\frictionfactor} \frac{8\density\gijEdge{\length}}{\gijEdge{\diameter}^5\pi^2}\abs{\gijtEdge{\flow}}\gijtEdge{\flow} \,,\quad \defSetPipes,
\end{equation}

The Darcy friction factor $\gijtEdge{\frictionfactor}$ is estimated using the Blasius correlation \cite{BlasiusH.1913}, which relates the friction factor to the Reynolds number $\Reynolds$.

\begin{equation}\label{eq:Blasius}
	\begin{aligned}
		\gijtEdge{\frictionfactor} &= 0.3164\left(\gijtEdge{\Reynolds}\right)^{-\frac{1}{4}}\,, \\
		\text{with}\quad\gijtEdge{\left(\Reynolds\right)} &= \frac{4 \density \gijtEdge{\abs{\flow}}}{\pi \viscosity \gijEdge{\diameter}} \quad \defSetPipes.
	\end{aligned}
\end{equation}

The heat loss of an insulated pipe installed underground is modeled the same as in Van der Heijde et al. \cite{van2017dynamic}. Consider $\gitNode{\dTinf}$ as the temperature difference at the node $\gi$ where the flow enters the pipe $\gi\gj$ and $\gijtEdge{\dTinf}$ at the pipe exit. The pipe exit temperature $\gijtEdge{\dTinf}$, due to heat loss to the environment, is then given by

\begin{equation}\label{eq:pipeEnergy}
	\gijtEdge{\dTinf} = \gitNode{\dTinf} \exp{\left(\frac{-\gijEdge{\length}}{\density \spHeatCap \abs{\gijtEdge{\flow}}{\gijEdge{(\thermR)}}}\right)} \quad \defSetPipes,
\end{equation}

where $\ve{\thermR}\in\Real{\npipes}$ are the thermal resistances per unit length of pipe between the water and the environment. For pipes with outer insulation jacket diameters $\ve{\diameter}_{\subScriptOuterD}\in\Real{\npipes}$ that are assumed to be larger than the inner diameters $\ve{\diameter}$ by a fixed ratio, i.e. $\ratioInsul=\frac{\ve{\diameter}_{\subScriptOuterD}}{\ve{\diameter}} \in\Real{} $, the combined thermal resistance of pipe and soil per unit length is \cite{Wallenten1991}
\begin{equation}
	\gijEdge{(\thermR)} = \frac{\ln(4 \depthPipe/(\ratioInsul \gijEdge{\diameter}))}{2 \pi \condGround} +\frac{\ln{\ratioInsul}}{2 \pi \condInsul}\,,\quad \defSetPipes,
\end{equation}
where $\condInsul\in\Real{}$ and $\condGround\in\Real{}$ are the thermal conductivities of the insulation and the surrounding ground, respectively, and $\depthPipe\in\Real{}$ is the depth at which the pipe is installed.

\subsection{Pipe junction model}
All nodes in the network represent pipe junctions, and for incompressible flow, conservation of mass is given by
\begin{equation} \label{eq:NodeContinutiy}
	\sum_{\inflowID} \flow_{\inflowID,\timeVar} - \sum_{\outflowID} \flow_{\outflowID,\timeVar} = 0\,,
\end{equation}
 and convected energy respectively

\begin{equation}\label{eq:NodeEnergy}
	\begin{aligned}
		& \sum_{\inflowID} \left(\max(\flow_\inflowID,0) \, \dTinf_\inflowID +\min(\flow_\inflowID,0) \,\dTinf_n\right)_{\timeVar}- \\& \sum_{\outflowID} \left(\max(\flow_\outflowID,0)\,\dTinf_n +\min(\flow_{\outflowID},0)\, \dTinf_{\outflowID}\right)_{\timeVar} = 0, \quad \forall n \in \setNodes\,,
	\end{aligned}
\end{equation}

where again $\flow_{\inflowID,\timeVar}$ with $\defInflow$ denotes the flow of incoming edges and $\ve{\flow}_{\outflowID,\timeVar}$ with $\defOutflow$ denotes the flow of outgoing edges of a node $n \in \setNodes$. The mixing of incoming flows is assumed to be perfect at the junction.
\subsection{Consumer model}

The flow in the consumer heating system is regulated by a control valve $\gijEdge{\radValve} \in \left[0,1\right],\defSetCon$. The momentum equation over this valve is 
\begin{equation} 
	\gijtEdge{\flow} = \gijtEdge{\radValve}\gijEdge{\valveRhydr} \sqrt{\gitNode{\pressure} - \gjtNode{\pressure}}\,, \quad \defSetRad,
\end{equation}
with $\gijEdge{\valveRhydr}$ a constant determined from nominal network operating conditions \cite{Pirouti2013}. The heat transfer in the consumer substation is modeled using the $\epsilon\text{-NTU}$ method, similar to the approach described by Guelpa et al. \cite{Guelpa2021}:
\newcommand{\heatHeatingSystem}[1]{\heat_{\rad#1}}
\begin{align}
	\density \spHeatCap \gijtEdge{\flow}(\giNode{\dTinf} - \gijEdge{\dTinf})_{\timeVar} - \gijtEdge{\left(\epsilonNTU \,\Cmin\right)}
	(\giNode{\dTinf}-\Tscold[,\gi\gj])_{\timeVar} =&\,0\,, \label{eq:HXConservationI}\\
	\gijtEdge{\left(\epsilonNTU \,\Cmin\right)}\left(\giNode{\dTinf}-\Tscold[,\gi\gj]\right)_{\timeVar} - \left[\left(\Tshot -\Tscold\right)\qs\right]_{\gi\gj,\timeVar}\density \spHeatCap =&\,0 \,, \label{eq:HXConservationII}\\
	\left[\left(\Tshot -\Tscold\right)\qs\right]_{\gi\gj,\timeVar}\density \spHeatCap - \heatHeatingSystem{,\gi\gj,\timeVar} =&\,0\,,\quad \defSetRad\,. \label{eq:HXConservationIII}
\end{align}

Here, $\gitNode{\dTinf}$ and $\gijtEdge{\dTinf}$ are the temperatures at the node inlet and the outlet on the primary side of the heat exchanger. $\veTshot[,\timeVar]\in\Real{\nRad}$ and $\veTscold[,\timeVar]\in\Real{\nRad}$ are the temperatures on the hot and cold side of the secondary side of the heat exchanger respectively, with $\veqs[,\timeVar]\in\Real{\nRad}$ being the flow rate in the heating system. The term $\ve{\epsilonNTU}_{\timeVar}\in\Real{\nRad}$ is the effectiveness of the heat exchanger, defined as a function of the number of transfer units $\ve{\NTU}_{\timeVar}\in\Real{\nRad}$ and the capacity ratio $\ve{\Cstar}\in\Real{\nRad}$ of the minimum heat capacity rate $\veCmin[,\timeVar]\in\Real{\nRad}$ to the maximum heat capacity rate $\veCmax[,\timeVar]\in\Real{\nRad}$: 

\begin{align}\label{eq:epsilonNTU}
	\gijtEdge{\epsilonNTU}&=\gijtEdge{\left(\frac{1-\exp(-\NTU(1-\Cstar))}{1-\Cstar\,\exp(-\NTU(1-\Cstar))}\right)}\,,\nonumber\\ 
	\gijtEdge{\NTU}&=\gijEdge{\left(\frac{\U\,\Ahexoriginal}{\Cmin[,\timeVar]}\right)}\,,\\	\gijtEdge{\Cstar}&=\gijtEdge{\left(\frac{\Cmin}{\Cmax}\right)}\,,\\
	\gijtEdge{(\Cmin)}&=\density\, \spHeatCap\left[\min(\flow,\qs)\right]_{\gi\gj,\timeVar}\,, \nonumber\\
	\gijtEdge{(\Cmax)}&=\density \spHeatCap\left[\max(\flow,\qs)\right]_{\gi\gj,\timeVar},\quad \defSetRad\,, \nonumber
\end{align}

with the heat transfer coefficient $\ve{\U}\in\Real{\nRad}$ and area $\ve{\Ahexoriginal}\in\Real{\nRad}$ of the heat exchanger. This system of equations is closed by defining $\veheatHeatingSystem{,\timeVar}$ with the characteristic radiator equation (see Wack et al. \cite{Wack2022a}) assuming knowledge of the consumers heating system.

In practice, the real design characteristics of the heat exchanger and heating system are often not known during the design stage. The $\ve{\U\Ahexoriginal}$ of the heat exchangers is assumed here to be selected based on the primary design temperatures $\veTphotdesign\in\Real{\nRad}$ and $\veTpcolddesign\in\Real{\nRad}$ and the heating system design temperatures $\veTshotdesign\in\Real{\nRad}$ and $\veTscolddesign\in\Real{\nRad}$. The design of a heat exchanger that meets the expected peak demand $\veQdemand[,{\peakPeriod}]$ is then obtained by

\begin{equation}
	\gijEdge{\left(\U\Ahexoriginal\right)} = \gijEdge{\left(\frac{\Qdemand[, {\peakPeriod}]}{\lmtd\left(\theta_{\mathrm{1h}}-\Tshot, \theta_{\mathrm{1c}}-\Tscold\right)_{\mathrm{nom}}}\right)}\,,
\end{equation}

$\defSetRad$, where $\veQdemand[,{\peakPeriod}]\in\Real{\nRad}$ is the peak heat demand of a consumer. Since the heating network operator usually has no direct control over the operation of the consumer heating system, the flow rate in the consumer heating system $\veqs[,\timeVar]$ is modeled as: 
\begin{equation} \label{eq:cons:qs}
	\gijtEdge{\left(\qs\right)}=\frac{1}{\density\spHeatCap}\gijEdge{\left(\frac{\Qdemand[,\timeVar]}{\Tshotdesign-\Tscolddesign}\right)}\quad \defSetRad\,,
\end{equation}
assuming that the flow rate in the consumer's heating system is adjusted according to demand variations.

\subsection{Producer model}
In the producer edges, a fixed input flow $\ve{\prodInput}_{\timeVar}\in\Real{\nPro}$ is imposed as boundary condition for this system of equations. In addition, a given temperature $\ve{\prodTemp}_{\timeVar}\in\Real{\nPro}$ is imposed for the heat source. This leads to 
\begin{equation} \label{eq:ProducerFixedTempFlow}
	\gijtEdge{\flow} = \prodInputi{\gi\gj,\timeVar},\qquad \gijtEdge{\dTinf} = \prodTempi{\gi\gj,\timeVar} \quad \forall \gi\gj \in 
	\Epro.
\end{equation}
To uniquely define the pressures throughout the network, a reference pressure is imposed in one of the producer return nodes.

\subsection*{Additional state constraints}\label{app:stateConst}
To ensure that the heat demand $\veQdemand[,\timeVar]\in\Real{\nRad}$ of every consumer is met, the following constraint is defined:
\begin{align}\label{eq:heatDemandLowerBound}
	-\frac{\heatHeatingSystem{,\gi\gj,\timeVar} - \Qdemandi{\gi\gj,\timeVar}}{\Qdemandi{\gi\gj,\timeVar}} \leq 0  \quad \defSetRad\,.
\end{align}
Here $\veheatHeatingSystem{,\timeVar}$ is the heat delivered to the building.

In addition to ensuring the structural integrity of the network components, a maximum pressure constraint is imposed:

\begin{align}\label{eq:maxPressure}
	\left(\gjtNode{\pressure}-\gitNode{\pressure}\right) - \maxPressure \leq 0 \quad \forall \gi\gj \in \Epro\,,
\end{align}

where $\maxPressure\in\Real{}$ represents the maximum allowable pressure difference in the network. 

\section{The adjoint gradient in a multi-period framework}\label{app:adjoint}
\newcommand{\deriv}[2]{\frac{\partial #1}{\partial#2}}
\newcommand{\mAdAdjVar}{\ve{\stateVarSkal^{*}}_{\timeVar}}
\newcommand{\mAdAdjVarInvariant}{\ve{\stateVarSkal^{*}}}
\newcommand{\mAdCost}{\costFull}
\newcommand{\mAdEqCon}{\equalCon_{\timeVar}}
\newcommand{\mAdStateVar}{\ve{\stateVarSkal}_{\timeVar}}
\newcommand{\mAdDesVarSteady}{\ve{\phi}}
\newcommand{\mAdDesVarTime}{\ve{\psi}}
\newcommand{\mAdDesVarTimeFull}{\designVar}
\newcommand{\mAdDesVar}{\designVar}
\newcommand{\forallt}{\quad\forall\timeVar\in\setPeriods}

Due to the quasi steady state assumption of the \gls{dhn} model, the evaluation of the adjoint gradients can be decomposed in time by redefining the cost function in equation (\ref{eq:totalCost}) as\footnote{This can be done because the capacity cost is time invariant and $\sum_{\timeVar=1}^{\timeSlices}\periodWeight_{\timeVar} = 1$}: 
\begin{equation}
	\begin{aligned} 
		\costFull\left(\designVar,\stateVar\right) &=\sum_{\timeVar\in\setPeriods} \periodWeight_{\timeVar}\costFull_{\timeVar}\left(\designVar_{\timeVar},\stateVar_{\timeVar}\right)\,.
	\end{aligned}
\end{equation}

This allows to compute the adjoint equations individually for every time period $\timeVar$ 

\begin{equation}
	\tp{\left(\deriv{\mAdEqCon}{\mAdStateVar}\right)}\mAdAdjVar = -\tp{\left(\deriv{\mAdCost}{\mAdStateVar}\right)}\quad\forallt\,, \label{eq:adjointEquationMP}
\end{equation}
further allowing to efficiently calculate the adjoint gradients individually
\begin{equation}
	\nabla\mAdCost_{\timeVar} = \periodWeight_{\timeVar}\tp{\left(\deriv{\mAdCost_{\timeVar}}{\mAdDesVar_{\timeVar}}\right)} + \tp{\left(\deriv{\mAdEqCon}{\mAdDesVar_{\timeVar}}\right)}\mAdAdjVar\,. \label{eq:adjointGradientMP}
\end{equation}

and later assemble them to the full gradient
\begin{equation}
	\begin{aligned}
		\nabla\mAdCost &= \sum_{\timeVar\in\setPeriods}\tp{\deriv{\mAdDesVar_{\timeVar}}{\mAdDesVar}}\left[\periodWeight_{\timeVar}\tp{\left(\deriv{\mAdCost_{\timeVar}}{\mAdDesVar_{\timeVar}}\right)} + \tp{\left(\deriv{\mAdEqCon}{\mAdDesVar_{\timeVar}}\right)}\mAdAdjVar\right]\\&= \sum_{\timeVar\in\setPeriods}\tp{\deriv{\mAdDesVar_{\timeVar}}{\mAdDesVar}}\nabla\mAdCost_{\timeVar}\,.
	\end{aligned}
\end{equation}

This allows the gradient calculation to be parallelized over all time periods. 

\section{The influence of the number of representative periods}\label{app:validation}
\newcommand{\error}{\varepsilon}
Aggregating the time series for both the heat demand and the outdoor temperature introduces an error in the cost of the optimized \gls{dhn} design. In an attempt to validate that the optimized design is independent of the temporal resolution at the chosen number of representative periods, this error is further investigated. Due to computational limitations, the highest number of periods considered in this study is $\nperiods=4$. Now the relative error of e.g. the total project cost $\costFull$ against the highest possible time resolution is calculated. Assuming that $\costFull_{\setPeriods_2}$ is the total project cost considering two time periods where $\setPeriods_2=\{1,\peakPeriod\}$, then the relative difference $\error$ against the project cost $\costFull_{\setPeriods_4}$ considering four time periods where $\setPeriods_4=\{1,2,3,\peakPeriod\}$ is calculated as follows:

\begin{equation}
	\error = \left|\frac{\costFull_{\setPeriods_2}-\costFull_{\setPeriods_4}}{\costFull_{\setPeriods_4}}\right|\,.
\end{equation}

\begin{figure}[ht]
	\centering
	\includegraphics{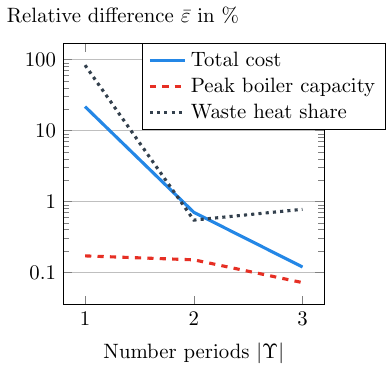}
	\caption{Evolution of the relative difference in total cost $\costFull$, peak boiler capacity and waste heat share of the optimized \gls{dhn} design for a different number of representative periods against the highest time resolution of $\card{\setPeriods}=4$. To ensure comparability, the performance of each design was evaluated at the highest time resolution.}
	\label{fig:numberPeriods}
\end{figure}

Now, the design and topology of the \gls{dhn} case of section \ref{sec:MPworstvsMulti} considering only one waste heat source and no producer unavailability is optimized for different temporal resolutions $\nperiods=1$, 2 \& 3. The relative difference of total cost, peak boiler capacity, and waste heat fraction against a temporal resolution of $\nperiods=4$ is visualized in figure \ref{fig:numberPeriods}. Here, the cost and performance of each optimized network design were evaluated at the highest temporal resolution of $\nperiods=4$ to ensure comparability. It should be noted that the relative difference in total cost quickly drops from $\SI{21.74}{\percent}$ for the worst-case design where $\setPeriods=\peakPeriod$ to $\SI{0.12}{\percent}$ for an optimized design considering three time periods with $\setPeriods=\{1,2,\peakPeriod\}$.

The rapid stagnation of the total project cost seems to indicate that the chosen temporal resolution of $\card{\setPeriods}=4$, including three representative periods and one worst case period, may be a good approximation of the full time series for the considered optimization problem and case study. It should be noted that the number of required representative periods is case-dependent.

\bibliography{modified_library,websites}

\end{document}